\input jytex.tex   
\typesize=10pt \magnification=1200 \baselineskip17truept
\footnotenumstyle{arabic} \hsize=6truein\vsize=8.5truein
\sectionnumstyle{blank}
\chapternumstyle{blank}
\chapternum=1
\sectionnum=1
\pagenum=0

\def\begintitle{\pagenumstyle{blank}\parindent=0pt
\begin{narrow}[0.4in]}
\def\endtitle{\end{narrow}\newpage\pagenumstyle{arabic}}


\def\beginexercise{\vskip 20truept\parindent=0pt\begin{narrow}[10
truept]}
\def\endexercise{\vskip 10truept\end{narrow}}


\def\eql#1{\eqno\eqnlabel{#1}}
\def\ref{\reference}
\def\peq{\puteqn}
\def\pref{\putref}

\def\mgn{\marginnote}
\def\bex{\begin{exercise}}
\def\eex{\end{exercise}}


\font\open=msbm10 

\font\goth=eufm10  

\def\StretchRtArr#1{{\count255=0\loop\relbar\joinrel\advance\count255 by1
\ifnum\count255<#1\repeat\rightarrow}}
\def\StretchLtArr#1{\,{\leftarrow\!\!\count255=0\loop\relbar
\joinrel\advance\count255 by1\ifnum\count255<#1\repeat}}

\def\StretchLRtArr#1{\,{\leftarrow\!\!\count255=0\loop\relbar\joinrel\advance
\count255 by1\ifnum\count255<#1\repeat\rightarrow\,\,}}

\def\mbox#1{{\leavevmode\hbox{#1}}}

\def\hspace#1{{\phantom{\mbox#1}}}

\def\oC{\mbox{\open\char67}}

\def\goa{\mbox{{\goth\char97}}}
\def\gob{\mbox{{\goth\char98}}}

\def\gl{\mbox{{\goth\char108}}}

\def\gs{\mbox{{\goth\char115}}}

\def\al{\alpha}
\def\bal{{\bmit\alpha}} 
\def\bbe{{\bmit\beta}} 
\def\brho{{\bmit\rho}} 
\def\bpsi{{\bmit\psi}} 
\def\bsi{{\bmit\sigma}} 
\def\bxi{{\bmit\xi}}
\def\be{\beta}
\def\ga{\gamma}
\def\de{\delta}

\def\ka{\kappa}
\def\la{\lambda}
\def\La{\Lambda}

\def\Om{\Omega}

\def\si{\sigma}
\def\Si{\Sigma}
\def\th{\theta}

\def\ze{\zeta}

\def\De{\Delta}

\def\caD{{\cal D}}

\def\caH{{\cal H}}

\def\Real{{\rm Re\,}}

\def\Imag{{\rm Im\,}}


\def\frac#1/#2{\leavevmode\kern.1em
\raise.5ex\hbox{\the\scriptfont0 #1}\kern-.1em/\kern-.15em
\lower.25ex\hbox{\the\scriptfont0 #2}}
\def\sfrac#1/#2{\leavevmode\kern.1em
\raise.5ex\hbox{\the\scriptscriptfont0 #1}\kern-.1em/\kern-.15em
\lower.25ex\hbox{\the\scriptscriptfont0 #2}}
\def\half{{1\over 2}}
\def\gtorder{\mathrel{\raise.3ex\hbox{$>$}\mkern-14mu
             \lower0.6ex\hbox{$\sim$}}}
\def\ltorder{\mathrel{\raise.3ex\hbox{$<$}\mkern-14mu
             \lower0.6ex\hbox{$\sim$}}}

\def\semidirprod{\rlap{\ss C}\raise1pt\hbox{$\mkern.75mu\times$}}
\def\for{\lower6pt\hbox{$\Big|$}}
\def\fish{\kern-.25em{\phantom{abcde}\over \phantom{abcde}}\kern-.25em}


\def\boxit#1{\vbox{\hrule\hbox{\vrule\kern3pt
        \vbox{\kern3pt#1\kern3pt}\kern3pt\vrule}\hrule}}
\def\dalemb#1#2{{\vbox{\hrule height .#2pt
        \hbox{\vrule width.#2pt height#1pt \kern#1pt \vrule
                width.#2pt} \hrule height.#2pt}}}

\def\ol{\overline}
\def\frac#1#2{{{#1}\over{#2}}}

\def\noin{\noindent}

\def\comb#1#2{{\left(#1\atop#2\right)}}

\def\etc{{\it etc. }}

\def\eg{{\it e.g.}}
\def\ie{{\it i.e. }}
\def\cf{{\it cf }}
\def\pa{\partial}

\def\ket#1{\mid#1\rangle}
\def\me#1#2#3{\langle{#1}\mid\!{#2}\!\mid{#3}\rangle}  

\def\Tr{{\rm Tr\,}}

\def\Threej#1#2#3#4#5#6{\biggl({#1\atop#4}{#2\atop#5}{#3\atop#6}\biggr)}
\def\threej#1#2#3#4#5#6{\bigl({#1\atop#4}{#2\atop#5}{#3\atop#6}\bigr)}
\def\3j#1#2#3#4#5#6{\left\lgroup\matrix{#1&#2&#3\cr#4&#5&#6\cr}
\right\rgroup}

\def\m?{\mgn{?}}

\def\pa{\partial}

\def\beq{\begin{eqnarray}}
\def\eeq{\end{eqnarray}}


\def\aop#1#2#3{{\it Ann. Phys.} {\bf {#1}} ({#2}) #3}
\def\ajop#1#2#3{{\it Am. J. Phys.} {\bf {#1}} ({#2}) #3}

\def\cmp#1#2#3{{\it Comm. Math. Phys.} {\bf {#1}} ({#2}) #3}
\def\cqg#1#2#3{{\it Class. Quant. Grav.} {\bf {#1}} ({#2}) #3}

\def\ijmp#1#2#3{{\it Int. J. Mod. Phys.} {\bf {#1}} ({#2}) #3}

\def\jmp#1#2#3{{\it J. Math. Phys.} {\bf {#1}} ({#2}) #3}
\def\jpa#1#2#3{{\it J. Phys.} {\bf A{#1}} ({#2}) #3}

\def\np#1#2#3{{\it Nucl. Phys.} {\bf B{#1}} ({#2}) #3}
\def\pl#1#2#3{{\it Phys. Lett.} {\bf {#1}} ({#2}) #3}

\def\prp#1#2#3{{\it Phys. Rep.} {\bf {#1}} ({#2}) #3}
\def\pr#1#2#3{{\it Phys. Rev.} {\bf {#1}} ({#2}) #3}
\def\prA#1#2#3{{\it Phys. Rev.} {\bf A{#1}} ({#2}) #3}

\def\prD#1#2#3{{\it Phys. Rev.} {\bf D{#1}} ({#2}) #3}

\def\rmp#1#2#3{{\it Rev. Mod. Phys.} {\bf {#1}} ({#2}) #3}

\def\zfp#1#2#3{{\it Z. f. Phys.} {\bf {#1}} ({#2}) #3}

\def\cras#1#2#3{{\it Comptes Rend. Acad. Sci. (Paris)} {\bf{#1}} (#2) #3}
\def\prs#1#2#3{{\it Proc. Roy. Soc.} {\bf A{#1}} ({#2}) #3}
\def\pcps#1#2#3{{\it Proc. Camb. Phil. Soc.} {\bf{#1}} ({#2}) #3}
\def\mpcps#1#2#3{{\it Math. Proc. Camb. Phil. Soc.} {\bf{#1}} ({#2}) #3}

\def\amsh#1#2#3{{\it Abh. Math. Sem. Ham.} {\bf {#1}} ({#2}) #3}
\def\am#1#2#3{{\it Acta Mathematica} {\bf {#1}} ({#2}) #3}
\def\aim#1#2#3{{\it Adv. in Math.} {\bf {#1}} ({#2}) #3}
\def\ajm#1#2#3{{\it Am. J. Math.} {\bf {#1}} ({#2}) #3}

\def\aom#1#2#3{{\it Ann. of Math.} {\bf {#1}} ({#2}) #3}
\def\cjm#1#2#3{{\it Can. J. Math.} {\bf {#1}} ({#2}) #3}
\def\bams#1#2#3{{\it Bull.Am.Math.Soc.} {\bf {#1}} ({#2}) #3}

\def\cmh#1#2#3{{\it Comm. Math. Helv.} {\bf {#1}} ({#2}) #3}

\def\dmj#1#2#3{{\it Duke Math. J.} {\bf {#1}} ({#2}) #3}
\def\invm#1#2#3{{\it Invent. Math.} {\bf {#1}} ({#2}) #3}

\def\jdg#1#2#3{{\it J. Diff. Geom.} {\bf {#1}} ({#2}) #3}

\def\joa#1#2#3{{\it J. of Algebra} {\bf {#1}} ({#2}) #3}
\def\jram#1#2#3{{\it J. f. reine u. Angew. Math.} {\bf {#1}} ({#2}) #3}
\def\jims#1#2#3{{\it J. Indian. Math. Soc.} {\bf {#1}} ({#2}) #3}
\def\jlms#1#2#3{{\it J. Lond. Math. Soc.} {\bf {#1}} ({#2}) #3}
\def\jmpa#1#2#3{{\it J. Math. Pures. Appl.} {\bf {#1}} ({#2}) #3}
\def\ma#1#2#3{{\it Math. Ann.} {\bf {#1}} ({#2}) #3}

\def\mz#1#2#3{{\it Math. Zeit.} {\bf {#1}} ({#2}) #3}
\def\ojm#1#2#3{{\it Osaka J.Math.} {\bf {#1}} ({#2}) #3}

\def\pems#1#2#3{{\it Proc. Edin. Math. Soc.} {\bf {#1}} ({#2}) #3}

\def\plb#1#2#3{{\it Phys. Letts.} {\bf {B#1}} ({#2}) #3}
\def\plms#1#2#3{{\it Proc. Lond. Math. Soc.} {\bf {#1}} ({#2}) #3}
\def\pgma#1#2#3{{\it Proc. Glasgow Math. Ass.} {\bf {#1}} ({#2}) #3}
\def\qjm#1#2#3{{\it Quart. J. Math.} {\bf {#1}} ({#2}) #3}

\def\rmjm#1#2#3{{\it Rocky Mountain J. Math.} {\bf {#1}} ({#2}) #3}

\def\tams#1#2#3{{\it Trans.Am.Math.Soc.} {\bf {#1}} ({#2}) #3}

\begin{title}
\vglue 1truein
\vskip15truept
\centertext {\Bigfonts \bf Spherical harmonics, invariant theory}
\vskip10truept \centertext{\Bigfonts \bf and Maxwell's poles}
 \vskip 20truept
\centertext{J.S.Dowker\footnote{dowker@man.ac.uk}} \vskip 7truept
\centertext{\it Theory Group,} \centertext{\it School of Physics and
Astronomy,} \centertext{\it The University of Manchester,}
\centertext{\it Manchester, England} \vskip40truept
\begin{narrow}
I discuss the relation between harmonic polynomials and invariant theory
and show that homogeneous, harmonic polynomials correspond to ternary
forms that are apolar to a base conic (the absolute). The calculation of
Schlesinger that replaces such a form by a polarised binary form is
reviewed.

It is suggested that Sylvester's theorem on the uniqueness of Maxwell's
pole expression for harmonics is renamed the Clebsch--Sylvester theorem.

The relation between certain constructs in invariant theory and angular
momentum theory is enlarged upon and I resurrect the Joos--Weinberg
matrices.

Hilbert's projection operators are considered and their generalisations
by Story and Elliott are related to similar, more recent constructions in
group theory and quantum mechanics, the ternary case being equivalent to
SU(3).

\end{narrow}
\vskip 5truept
\vskip 60truept
\vfil
\end{title}
\pagenum=0
\newpage

\section{\bf 1. Introduction.}
There has recently been increased activity in some aspects of classical
spherical harmonic theory partly in response to the very accurate
measurements of the cosmic microwave background and the need to analyse
these in a significant fashion. One approach has lead to the rediscovery
of Maxwell's way of picturing spherical harmonics through their `poles',
(\eg\ [\pref{CHS,KandW,Weeks1,Dennis,SSJR}]), the uniqueness of which
description is the content of Sylvester's theorem.

Because of the relation between this topic, and angular momentum theory
in general, and invariant theory, I thought it might be helpful to set
down a personal resum\'e of some of these things in a more historical
context, if only to draw attention to sometimes forgotten, but relevant,
work. I therefore try to draw together mostly existing material. I also
include some peripheral constructions that I feel are worth exposing
again, such as the Joos--Weinberg matrices.

The heyday of classical, constructive invariant theory was the second
half of the 19th century and the first quarter of the 20th. However, it
never really went away and has recently undergone a resurgence of
interest (\eg\ Olver, [\pref{Olver}]) finding applications in coding,
computer aided design and pattern recognition.\footnote{ Olver's
introduction in [\pref{Olver}] and Sturmfels' in Hilbert,
[\pref{Hilbert2}], supply useful historical comments.}

It is possible to find parallels (and equivalences) between many
constructions and processes in classical invariant theory and those in
`modern' applied mathematics. Although my pointing out some of these is
probably more of interest than utility, I hope to provide something of
value.

\section{\bf 2. Spherical harmonics.}
Discussions of spherical harmonics are many. The term itself seems to
date to the treatise of 1867 by Thomson and Tait, [{\pref{TandT}], where
fairly general definitions can be found, but the subject had its
beginning in a work of Legendre of 1782 followed by Laplace's paper of
1782 on potential theory (as we now call it). Later fundamental papers
are those of Green (1828) and Gauss (1841). Maxwell, [\pref{Maxwell}],
lists some standard technical references of his time.

As might be expected, there are a number of approaches. Here I just try
to give some relevant facts without bothering too much about logical
ordering.

For this reason, I begin with the addition theorem for (solid) spherical
harmonics which can be regarded as `classic', being derived, effectively,
by Legendre. It is,
  $$
  (r'\,r)^L
   P_L(\cos\ga)=C^L_M({\bf r'})\,C^M_L({\bf r})
  \eql{addf}
  $$
where $\ga$ is the angle between ${\bf r}$ and ${\bf r'}$, $ r\,r'
\cos\ga={\bf r\,.\,r'}\,. $

I give here, because it's convenient, the useful composition law for
these standard solid harmonics, $C_L^M({\bf r})$, in my conventions,
  $$
  C_{L_1}^{M_1}({\bf r})\,C_{L_2}^{M_2}({\bf r})=
 -\sum_{L_3}(ir)^{L_1+L_2-L_3}\Threej{L_1}{L_2}{L_3}000
 \Threej{M_1}{M_2}{L_3}{L_1}{L_2}{M_3}\,C_{L_3}^{M_3}({\bf  r})\,,
  \eql{ad1}
  $$
and the `inverse',
  $$
  C_{L_1}^{M_1}({\bf r})\,C_{L_2}^{M_2}({\bf r})
  \Threej{L_1}{L_2}{M_3}{M_1}{M_2}{L_3} =
 -(ir)^{L_1+L_2-L_3}\Threej{L_1}{L_2}{L_3}000C_{L_3}^{M_3}({\bf
  r})\,.
  \eql{ad2}
  $$

The left--hand side of (\peq{addf}) is a homogeneous bipolynomial, of
degree $2L$, in ${\bf r}$ and ${\bf r'}$ and is homogeneous of degree $L$
in ${\bf r}$ and of degree $L$ in ${\bf r'}$. It is symmetrical in ${\bf
r}$ and ${\bf r'}$ and is a solid spherical harmonic in either set being
sometimes referred to as the {\it biaxal} harmonic of ${\bf r}$ and ${\bf
r'}$, ( [\pref{TandT}] \footnote{ Thomson and Tait, [\pref{TandT}] p.159,
derive (\peq{addf}) by means of Taylor's theorem. See also Hobson,
[\pref{Hobson}] \S 90.}, [\pref{Hobson}] ). $P_L(\cos\ga)$ is a Laplace
coefficient (\cf\ MacMillan, [\pref{MacMillan}] p.377 \footnote {As noted
by MacMillan on p.300, (\peq{addf}) can be thought of as a transformation
from a harmonic with its pole on the $z$--axis to one whose pole is on
the line $(x',y',z')$.}) and the explicit form of the left--hand side of
(\peq{addf}) can be found from the classic series for $P_L$ as, (\cf\
[\pref{MacMillan}] p.383 equn.(2)),
  $$\eqalign{
   (r'\,r)^L\,& P_L(\cos\ga)\equiv H_L({\bf r},{\bf r'}))\cr
   &={1\over 2^L}\sum_K(-1)^K\comb LK \comb{2L-2K}L\,
   r'^{\,2K}\,r^{2K}({\bf r\,.\,r'})^{L-2K}\,. }
  \eql{biaxal}
  $$

Conversely one could derive this expression from first principles, using
rotational invariance and the harmonic condition, and then deduce the
standard series expansion of $P_L$, \cf\ [\pref{BandL}].

The vectors ${\bf r}$ and ${\bf r'}$ can be taken complex, subject to
complex orthogonal transformations (complex rotations or, equivalently,
Lorentz transformations). An important case is when one of them, say
${\bf r'}$, is isotropic, or null, $r'=0$, which can be achieved by
treating the spin--one quantity ${\bf r'},\equiv {\goa},$ as composed
from a 2--spinor, $\psi=\big({\xi\atop\eta}\big)$ $\in\oC^2$,
    $$
    \goa_1= -{\goa_x-i\goa_y\over i\sqrt2}=\xi^2\,\,,\quad \goa_{-1}=
    {\goa_x+i\goa_y\over i\sqrt2}=\eta^2\,
    \,,\quad\,\goa_0=-i\goa_z=\sqrt2\,\xi\eta\,,
   \eql{cm}
    $$
sometimes called the Cartan map, [\pref{BandL}].

$(\goa_x,\goa_y,\goa_z)$ are the Cartesian components of $\goa$ and I
have chosen a normalisation in keeping with the standard/contrastandard
definitions. This accounts for the differences with [\pref{BandL}]. I
should therefore say that my convention is that upstairs indices
correspond to contrastandard, and downstairs ones to standard behaviour,
in the terminology of Fano and Racah [\pref{FandR}], to whom I generally
adhere for signs, factors of $i$ \etc However my $3j$--symbols are the
more usual ones of Wigner and my angle and rotation conventions follow
those of Brink and Satchler, [\pref{BandS}]. The relation between their
surface harmonics and my solid ones is
  $$
  r^L\,C^{(BS)}_{LM}(\th,\phi)=(-i)^L\, C_L^M({\bf r})\,.
  $$
I also give my raising and lowering convention, which is the transpose of
Wigner's and is the same as that adopted by Williams, [\pref{Williams}].
Specifically, for a spin--$j$ object, $\phi$, I set
  $$ \phi^m=(-1)^{j+m}\phi_{-m}\,.
  \eql{raising}
  $$

Using (\peq{cm}), (\peq{addf}) reduces to,
   $$
   {(2L)!\over2^L L!^2}\,(\goa\,.\,{\bf r})^L=
   C^L_M(\goa)\,C^M_L({\bf r})\,.
   \eql{cmn}
   $$
$C_M^L(\goa)$ has the familiar, monomial expression (\eg\ Weyl,
[\pref{Weylqm}], van der Waerden, [\pref{Waerden}], Wigner,
[\pref{Wigner}], Bargmann, [\pref{Bargmann}], Schwinger,
[\pref{Schwinger}]),
   $$\eqalign{
   C_M^L(\goa)&={(2L)! \over 2^{L/2} L!}
   {\xi^{L+M}\eta^{L-M}\over\big[(L+M)!(L-M)!\big]^{1/2}}\cr
  &={\sqrt{(2L)!}\over2^{L/2} L!}\,\xi^{(L)}_M\,,
    }
   \eql{nsph}
   $$
where $\xi^{(\!j)}_m$ is (for $j$ half--integral) what I have termed,
[\pref{DowkandG}], a {\it null} $(2j+1)$--spinor,
  $$
  \xi^{(\!j)}_m= \comb{2j}{j-m}^{1/2}\,\,
  \xi^{j+m}\eta^{j-m}\,.
  \eql{nspin}
  $$

The factors here\footnote{ Landau and Lifschitz, a very useful book, make
the same choice, [\pref{LandL}] eqn.(97.4). See also Fano and Racah,
[\pref{FandR}], App.F.} are chosen so that the scalar product of
$\xi^{(\!j)}_m$ with another null spinor,
  $$
  \al^{(\!j)}_m= \comb{2j}{j-m}^{1/2}\,\,\al^{j+m}\be^{j-m}\,,
  $$
is the simple power of a bracket, (\cf\ [\pref{BandL}] (6.150)),
  $$
 \al_{(\!j)}^m\, \xi^{(\!j)}_m=(\eta\al-\xi\beta)^{2j}=\left|
   \matrix{\al&\be\cr\xi&\eta}\right|^{2j}\,.
   \eql{spsp}
  $$

In terms of this notation, (\peq{cmn}) reads,
  $$
   {\sqrt{(2L)!}\over2^{L/2} L!}\,(\goa\,.\,{\bf r})^L=
   \xi^{(L)}_M\,C^M_L({\bf r})\,,
   \eql{cmn2}
   $$
quoted, with conventional phases, by Schwinger [\pref{Schwinger}]; see
[\pref{BandL}] 6.149. The left--hand side can be considered to be a
generating function for the spherical harmonics (\cf\ Erd\'elyi {\it et
al}, [\pref{EMOT}], \S 11.5.1). Likewise, and rather trivially, the
right--hand side of (\peq{spsp}) can be taken as a generating function
for the null spinor $\xi_{(j)}$. Just set $t\equiv \al/\be$ and compare
with [\pref{EMOT}] equn 11.7(10).

If ${\bf r}$ is also a null vector, ${\bf r}=\gob$, constructed from the
2--spinor $\big({\al\atop \be}\big)$ by the Cartan map, then, from
(\peq{spsp}) and (\peq{cmn2}), or directly from (\peq{cm}),
  $$
  (\goa\,.\,\gob)^L=(\eta\al-\xi\beta)^{2L}\,.
  \eql{nn}
  $$

As an interlude, I now rederive (\peq{cmn2}) in order to introduce some
related material that I feel deserves revisiting. I start from the
equation for the product of $2j$ null 3--vectors, (see [\pref{DowkandG}]
equn.(9)),
  $$
  \goa^{i_1}\ldots\goa^{i_{2j}}= \rho_j\,\xi^m_{(j)}\,\big({t^{i_1\ldots
  i_{2j}}}\big)_m^{\,\,\,\,n}\,\xi^{(j)}_n\equiv\rho_j\,\overline\xi\,
  {t^{i_1\ldots i_{2j}}}\,\xi\,,
  \eql{hscm}
  $$
where the ${t^{i_1\ldots i_{2j}}}$ are the spatial components of the
Joos--Weinberg matrices, $t^{(\mu)}= {t^{\mu_1\ldots
  \mu_{2j}}}$ which are higher--spin analogues of
the {\it space--time} Pauli matrices, $\si^\mu$. They are symmetric and
traceless on $({\mu})$. $\rho_j$ is the normalisation,
  $$
  \rho_j=2^{-2j}e^{-i\pi j}\,(4j)!^{1/2}\,.
  \eql{rho}
  $$

They form a complete set (another one!) of $(2j+1)\times(2j+1)$ hermitian
matrices and can be expressed as symmetrized powers of the angular
momentum matrices, examples being, (beware: I use `$j\,$' in two roles),
  $$
  t^{0\ldots0}={\bf 1};\quad t^{i0\ldots0}={1\over j}\,J^i;\quad
  t^{ij\,0\ldots0}={1\over2j-1}\bigg({1\over
j}[J^i,J^j\,]_+-\de^{ij}\bigg)\,.
  \eql{texamp}
  $$

The $t^{(\mu)}$, when sandwiched between $(2j+1)$--spinors, transform
like 4--tensors. Equation (\peq{hscm}) is a higher--spin version of the
Cartan map (\peq{cm}). The simplest case is $j=1/2$, identical to
(\peq{cm}),
  $$
  \goa_i={1\over i\sqrt2}\,\overline\psi\,\bsi_i\,\psi\,,\quad
  \psi=\left(\matrix{\xi\cr\eta}\right)\,,\quad
  \overline\psi=(-\eta,\xi)\,,
  \eql{pauli}
  $$
where $\bsi_i$ are the standard spatial Pauli matrices.

I now contract (\peq{hscm}) with ${\bf r}$s to give,
  $$
  (\goa\,.\,{\bf r})^{2j}=\rho_j\,\overline\xi\,( r_i r_j\ldots
  t^{ij\ldots})\,\xi\,.
  \eql{hscm2}
  $$

From the tensor operator (adjoint) transformation of the $t^{ij\ldots}$
one can easily show that, [\pref{DandD}],\mgn{check $i$s}
   $$
     r_i r_j\ldots t^{ij\ldots}=e^{i\pi j}\,r^{2j}\,
    e^{-i\pi\widehat{\bf r}.{\bf J}}\,.
   \eql{jwm}
   $$

Further, the important multiplication law for null spinors is,
  $$
  \xi^{(j_1)}_{m_1}\,\xi^{(j_2)}_{m_2}=(-1)^{2j_1}(2j_1+2j_2+1)^{1/2}
  \Threej {j_1}{j_2}{m_3}{m_1}{m_2}{j_1+j_2}\,   \xi^{(j_1+j_2)}_{m_3}\,,
  \eql{ncomb}
  $$
essentially by Wigner--Eckart. This means, in particular, that,
  $$
  \xi^{(j_1)}_{m_1}\,\xi^{(j_2)}_{m_2}
  \Threej {m_1}{m_2}{j_3}{j_1}{j_2}{m_3}=0\,,
  \eql{ncomb2}
  $$
unless $j_3=j_1+j_2$.

So, from (\peq{ncomb}), (\peq{hscm2}) reads,
   $$
    (\goa\,.\,{\bf r})^{2j}=\rho_j\,(4j+1)^{1/2}\,e^{i\pi
    j}\,(-r)^{2j}\,\xi_m^{(2j)}\,
    \Tr\big(u^m_{2j}(j)\,\caD^{(j)}(\pi\widehat{\bf r})\big)\,,
    \eql{hscm3}
   $$
where $u_L^M(J)$ are the matrices discussed by Racah and have matrix
elements,
  $$
 \big[u^M_L(J)\big]_m^{\cdot m'}=\me{Jm}{u_L^M({\bf J})}{Jm'}
 =\Threej JM{m'}mLJ\,.
 \eql{ewes}
 $$

The trace on the right--hand side of (\peq{hscm3}) is a hyperspherical
harmonic,\footnote{ These are standard objects and will form the basis of
a further work. I refer here now only to Talman, [\pref{Talman}].} which
is a spin--$j$ SU(2) representation matrix expressed in polar
coordinates, $(\chi,\xi,\eta)$, on the 3--sphere. It is evaluated on the
equator of S$^3$, $\chi=\pi/2$, and is just a surface harmonic, $\oC$.
Generally,
   $$
   \Tr(u^M_L\,\caD^J(g))=i^LH_{J,L}(\chi)\,\oC^M_L(\xi,\eta)\,,
   \eql{hspol}
   $$
where the radial function is related to Gegenbauer polynomials,
  $$
   H_{J,L}(\chi)=L!\bigg[{(2J-L)!\over(2J+L+1)!}\bigg]^{1/2}
  (2i\sin \chi)^L\, C^{L+1}_{2J-L}(\cos \chi)\,.
  \eql{hfun2}
  $$

In (\peq{hscm3}), $\chi=\pi/2$, $J=j$ and $L=2j$ so one needs,
  $$
  H_{j\,,\,2j}(\pi/2)=(2i)^{2j}\,(2j)!{1\over(4j+1)!^{1/2}}\,.
  \eql{spech}
  $$
Therefore, from (\peq{hspol}), using (\peq{spech}) and (\peq{rho}),
  $$\eqalign{
  (\goa\,.\,{\bf r})^{2j}&={\rho_j\over(4j)!^{1/2}}\,e^{i\pi j}\,
  (2i)^{2j}\,(-r)^{2j}\,
   \oC^m_{2j}(\widehat{\bf r})\,\xi_m^{(2j)}\cr
   &=2^{2j}{\rho_j\over(4j)!^{1/2}}\,
   C^m_{2j}({\bf r})\,\xi_m^{(2j)}\cr
  &= e^{-i\pi j}\,C^m_{2j}({\bf r})\,\xi_m^{(2j)}\,.
  }
  $$
I have thus regained (\peq{cmn2}), as promised, with integer $M=m$.

I wish to prove something that might not be immediately apparent \ie\
that the quantity on the right--hand side of (\peq{hscm}) is traceless,
as it ought to be since the $\goa$ are null. The spatial $t^{i\ldots j}$
themselves are not traceless but the null nature of $\xi$ saves the day
as I now show. Firstly, the space--time traceless property gives the
spatial trace,
  $$
  {t_i}^{i\,i_1\ldots i_{2j-2}}=t^{00\,i_1\ldots i_{2j-2}}\,,
  $$
which contains at most a product of $(2j-2)$ angular momentum matrices
and the claim, therefore, is that the quantity,
  $$
  \overline\xi^{(j)} J^{i_1}\ldots J^{i_n}\,\xi^{(j)}\,,\quad n\le2j-2\,,
   \eql{tracel}
  $$
vanishes. Writing the $J^i$ as $3j$ symbols, the products can be
recombined in turn using $6j$ symbols until the $\xi$s are coupled by the
same $3j$ symbol and the null theorem (\peq{ncomb2}) can be applied.
Because $j_3$ is never greater than $2j-2$, the desired result follows.
Alternatively, (\peq{ncomb}) can be applied directly to (\peq{tracel}) to
give,
  $$
  \bxi\,.\,\Tr\big({\bf u}_{2j}({\bf J})\, J^{i_1}\ldots J^{i_n}\big)\,,
  $$
in terms of the Racah $u$ matrix, (\peq{ewes}), which can be shown to
vanish. The ${\bf J}$ are spin--$j$ matrices. I leave the details as an
exercise.

I give some further details about the Joos--Weinberg matrices (Weinberg
[\pref{Weinberg}]) in particular on the relation (\peq{jwm}). In fact
this relation, or rather the one from which I shall shortly derive it,
was used by Weinberg to compute the matrices as products of angular
momentum matrices. A direct algebraic method, which is easily implemented
is the following. Setting,
  $$
  e^{-i\th{\bf n\,.\,J}}=\sum_{k=0}^{2j}c_k(\th)\,({\bf n\,.\,J})^k\,,
  $$
for $\th=\pi$, only even powers occur for integral $j$ and odd powers for
$j$ half--odd integral and one has to invert the set of equations,
obtained by diagonalisation,
  $$\eqalign{
   (-1)^m&=\sum_{k=0}^{j}m^{2k}c_{2k}(\pi),\,\quad m=1,\ldots j\,,
   \quad j\,\,\,{\rm integral}\cr
  i(-1)^{m-1/2}&=\sum_{k=0}^{j-1/2}m^{2k+1}c_{2k+1}(\pi),
   \,\quad m=1/2,\ldots j\,,\quad j\,\,\,{\rm half-odd\,\, integral}\,,
  }
  $$
with $c_0(\pi)=1$. Torruella, [\pref{Torruella}], provides an analytic
inversion as a function of $j$ but it is easier (and equivalent), for any
particular $j$, to use the tabled formulae,
  $$\eqalign{
  (-1)^m&=\sum_{n=0}^{j}(-1)^n2^{2n}{m^2\big(m^2-1^2\big)\big(m^2-2^2\big)
  \ldots\big(m^2-(n-1)^2\big)\over(2n)!}\cr
  (-1)^{m-1/2}&=\!\!\sum_{n=0}^{j-1/2}\!\!(-1)^{n}2^{2n+1}
  {m\big(m^2\!-(1/2)^2\big)\big(m^2\!-(3/2\big)^2)
  \ldots\big(m^2-(n-1/2)^2\big)\over(2n+1)!}\,.
  }
  $$

For example,
  $$\eqalign{
  j&=1,\qquad c_2(\pi)=-2\cr j&={3\over2},\qquad c_1(\pi)
  =7i/3,\,\,\,c_3(\pi)=-4i/3\cr
  j&=2,\qquad c_2(\pi)=-8/3,\,\,\,c_4(\pi)=2/3\,.
  }
  $$

From the computed expansion, using (\peq{jwm}), the spatial
Joos--Weinberg matrices  can be extracted as sums of symmetrised products
of the ${\bf J}^i$, if desired. I will not do so here but refer to the
calculation in Weinberg's paper, of which the above is a particular case.
(Just set $q_0=0$ in that reference.)

The $t^{(\mu)}$ are actually geared to the relativistic situation. They
transform under Lorentz transformations, regarded conveniently as complex
3--rotations, as
  $$
   e^{(\bal-i\bbe)\,.\,{\bf J}}\,\,t^{\mu\ldots}\,\,e^{(\bal+i\bbe)\,.\,{\bf J}}=
  t^{\nu\ldots}\,\,\La_\nu^{\,\,\,\mu}\ldots
  \eql{tlt}
  $$
where $\La$ is the 4--vector Lorentz transformation. The first thing to
note is that $t^{00\ldots0}$ is a scalar under the spatial rotation
subgroup, $\bal=0$, and so can be set equal to ${\bf 1}$ by Schur's
lemma. With this in mind, the next step, [\pref{Weinberg}], is to
saturate the $\mu$ indices with the four--vector, $q_\mu$, and choose
$\La$ to be a pure Lorentz boost, $\bbe=0$, that kills the 3--vector
part, \ie\ $q_\mu \to q_\mu'=(q,{\bf 0})$, where $q^2=q_0^2-|{\bf q}|^2$.
Then (\peq{tlt}) becomes,
  $$
  q_\mu\;\dots\,t^{\mu\ldots}=q^{2j}\,e^{-2\al\hat q\,.\,{\bf J}}\,,
  \eql{jwm2}
  $$
where
  $$
     \hat q= {{\bf q}\over|{\bf q}|},\quad \sinh\al={|{\bf q}|\over q}\,.
  $$
Expansion of the right--hand side of (\peq{jwm2}), as a polynomial in
$q_\mu$, allows the $t^{(\mu)}$ to be read off, as mentioned before,
[\pref{Weinberg}].

To obtain (\peq{jwm}) from (\peq{jwm2}), formally continue $q_0$ to zero.
Then $\sinh\al=i$ so $\al=i\pi/2$ and we are back to ordinary rotations
(through $\pi$!).

Before leaving this topic, I briefly mention two uses of the $t^{(\mu)}$.
Firstly, the massless equations (Weyl, Maxwell, linearised vacuum
gravity, Dirac's higher spin) can all be written in the form,
[\pref{Dowzilch}],
  $$
    t^{\mu\ldots\nu}\,\pa_\nu\,\phi(x)=0\,,
  $$
and, secondly, the quantity $\phi_{(j)}^{{\dag}}\,t^{(\mu)}\,\phi_{(j)}$
is a higher--spin analogue of the Bel--Robinson tensor, [\pref{DandD}].

\section{\bf 3. Invariant Theory.}

If $\xi$ and $\eta$ are viewed as {\it variables}, and $\al$ and $\be$ as
{\it coefficients}, the bracket power in (\peq{spsp}) is the symbolic
form of a binary $2j$--ic, traditionally written $(\al_1x_1+\al_2
x_2)^n$, $n=2j$ (\eg\ Grace and Young [\pref{GandY}]). For example, for
the {\it general} quadratic (spin one),  $a x_1^2+2b\,x_1\,x_2+cx_2^2$,
one has $a=\al_1^2$, $b=\al_1\al_2$ and $c=\al_2^2$, but $ac\ne b^2$ and
the coefficients, $(a,\sqrt2\,b,c)$, do not constitute the (spherical)
components of a {\it null} 3--vector (3--spinor),

Comparing with (\peq{cm}), it cannot escape notice that the Cartan map
had already appeared algebraically in the 1850's, when Aronhold
introduced the symbolic method into invariant theory. Since this theory
plays a part in what I wish to say, I present some expository material. I
will not avail myself of the various renderings of classical invariant
theory into modern algebraic terms, preferring the earlier language.

The general $2j$--ic, $\Phi_{2j}$, formed with the coefficients
$\phi_{(j)}^m$, is written,
  $$\eqalign{
  \Phi_{2j}(\ol\phi,\xi,\eta)&=\phi_{(j)}^m\,\xi_m^{(j)}\equiv \overline\phi\,\xi\cr
  &=(-1)^{2j}\xi^m_{(j)}
  \phi^{(j)}_m\equiv(-1)^{2j}\overline\xi\,\phi\,.
  }
  \eql{quantic}
  $$

In the case that the coefficients form a null $(2j+1)$--spinor, say
$\phi_{(j)}=\al_{(j)}$, $\Phi_{2j}$ takes the simple power form
(\peq{spsp}) and the $2j$ roots of $\Phi_{2j}=0$, for $t\equiv \xi/\eta$,
are all equal to $\al/\be$.\footnote{ This is an extreme example of a
{\it nullform}, as defined by Hilbert to be a form all of whose
invariants vanish. For this one needs only a root of multiplicity
$[n/2]+1$, $n$ being the form order.} The necessary and sufficient
condition for all roots to be equal, is that the Hessian of the quantic
vanish identically. The Hessian for the quantic (\peq{quantic}) is given
in my notation, by
  $$
  H=\Threej{2j-2}jj lmn\,\xi_{(2j-2)}^l\,\phi_{(j)}^m\,\phi_{(j)}^n\,,
  \eql{hessian}
  $$
and so one has the theorem, [\pref{DowkandG}], that if,
  $$
  \Threej{2j-2}jj lmn\,\phi_{(j)}^m\,\phi_{(j)}^n=0\,,
  $$
then $\phi_{(j)}$ is null. The converse is contained in (\peq{ncomb2}).

It is fundamental that subjecting the two spinor,
$\psi=({\xi\atop\eta})$, to a transformation belonging to SL(2,C), and
requiring the binary $n$--ic, $\Phi_n$, (\peq{quantic}), to be invariant,
induces a (particular) Lorentz transformation on the coefficients,
$\phi_m$. Invariant theory treatments can be found in Turnbull,
[\pref{Turnbull}], Chap.VIII \S8, Glenn, [\pref{Glenn}] \S1.2.4,
[\pref{GandY}], \S16.\footnote{ The group in invariant theory does not
have to be unimodular. For those requiring the basics of invariant theory
in short, but expert, compass I recommend Dickson's book,
[\pref{Dickson}], which treats both symbolic and non--symbolic methods
and employs group theory concepts.} I repeat some standard details here
out of interest.

The basic transformation $\psi \to\psi'$, expressed as the inverse for
convenience, is,
  $$
  \left(\matrix{\xi\cr\eta}\right)=\left(\matrix{l_1&m_1\cr l_2&m_2}\right)
   \left(\matrix{\xi'\cr\eta'}\right)\,.
  $$

The imposed invariance of $\Phi$ implies, setting $\xi'/\eta'=t$,
  $$\eqalign{
  \Phi_n(\ol\phi,\xi,\eta)&=\Phi_n(\ol\phi',\xi',\eta')=\eta'^n\,
  \Phi_n(\ol\phi',t,1)\cr
  &=\eta'^n\,
  \Phi_n(\ol\phi',l_1t+m_1,l_2t+m_2)\cr
  &=\eta'^n\,e^{t\,{\bf l.}\nabla_{\!\bf m}}\, \Phi_n(\ol\phi,m_1,m_2)
   }
  \eql{inv1}
  $$
where ${\bf l\,.}\nabla_{\!\bf m}$ is the {\it polarisation} operator (or
{\it translation} operator or directed derivative),
  $$
  {\bf l.}\nabla_{\!\bf m}\,=\,l_1{\pa\over\pa m_1}+l_2{\pa\over\pa m_2}
  \,\equiv\,\bigg(l{\pa\over\pa m}\bigg)\,.
  \eql{polop}
  $$

I interject a small, necessary calculational point. The normalisation of
the coefficients, $\phi^m$, differs from that of the coefficients as
usually defined in invariant theory. The two conventions are exhibited in
the forms,
  $$\eqalign{
  \Phi_{2j}(\ol\phi,\xi,\eta)&=\sum_{m=-j}^j\phi^m\, \comb{2j}{j-m}^{1/2}\,\xi^{j+m}\eta^{j-m}\cr
  &=\sum_{r=0}^na_r\,\comb nr\,\xi^{n-r}\eta^r=\Phi_n({\bf a},\xi,\eta)
  }
  \eql{coeffc}
  $$
with $n=2j$ and $r=j-m$, so that,
  $$
  \phi^{j-r}=\comb nr^{1/2}\,a_r\,,
  \eql{coeffs2}
  $$
connects the two sets of coefficients.

The normalisation for the $\phi$s corresponds to the use of the 1--$j$
symbol as the raising and lowering metric in weight space,
(\peq{raising}), as in (\peq{quantic}), and elsewhere, and this is an
appropriate place to make some elementary remarks on duality which I will
need later.

The spinor $\overline\psi=(-\eta,\xi)$ is dual to
$\psi=({\xi\atop\eta})$. I can make this projective by setting $-\eta\to
u$ and $\xi\to v$, and then refer to $u,v$ as line, or tangential,
coordinates. $\xi$ and $\eta$ are `point coordinates'.

The point binary form (\peq{coeffc}) can be written in line coordinates
as follows,
  $$\eqalign{
  \Phi_{2j}(\ol\phi,\xi,\eta)&=\comb{2j}{j-m}^{1/2}(-1)^{j-m}\,\phi^m \,\xi^{j+m}(-\eta)^{j-m}\cr
  &=(-1)^{2j}\comb{2j}{j-m}^{1/2}\,\,\phi_{-m}\, \,v^{j+m}u^{j-m}\cr
  &=(-1)^{2j}\,\comb{2j}{j-m}^{1/2}\,\phi_{m}\, \,u^{j+m}v^{j-m}\cr
  &=(-1)^{2j}\,\Phi_{2j}(\phi,u,v)\,,\cr
  }
  \eql{coeffd}
  $$
corresponding to ({\peq{quantic}). The sign factor could be absorbed into
the coefficients of the line form.

Putting (\peq{coeffc}) into the invariance (\peq{inv1}) yields
  $$
  \sum_{r=0}^n\comb nr a'_r\,t^{n-r}
  =e^{t\,{\bf l.}\nabla_{\!\bf m}}\, \Phi_n({\bf a},m_1,m_2)
  $$
and expansion gives the relation between the old and new coefficients,
  $$
  a'_r={r!\over n!}\,\big({\bf l\,.}\nabla_{\bf m}\big)^{n-r}\,
  \Phi_n({\bf a},m_1,m_2)\,,
  \eql{indt1}
  $$
\ie
  $$
  \phi'^m=\sqrt{{(j-m)!\over (2j)!(j+m)!}}
  \,\big({\bf l\,.}\nabla_{\bf m}\big)^{j+m}
  \,\Phi_{2j}(\ol\phi,m_1,m_2)\,.
  \eql{indt3}
  $$

If one rather uses $\eta'=t\xi'$, then the alternative formula,
  $$
  a'_r={(n-r)!\over n!}\,\big({\bf m\,.}
  \nabla_{\bf l}\big)^r\,\Phi_n({\bf a},l_1,l_2)\,,
  \eql{indt2}
  $$
\ie
  $$
  \phi'^m=\sqrt{{(j+m)!\over
  (2j)!(j-m)!}}\,\big({\bf m\,.}\nabla_{\bf l}
  \big)^{j-m}\,\Phi_{2j}(\ol\phi,l_1,l_2)\,,
  \eql{indt4}
  $$
arises ([\pref{Turnbull}], [\pref{Glenn}]).
\section{\it Infinitesimal behaviour and projection operators.}
As mentioned, the induced transformation on the coefficients is that of
the spin--$j$ representation of the rotation group (when SL(2,C) is
reduced, inessentially,  to SU(2)) and these relations are more or less
identical to a standard method of finding the representation matrices,
$\caD^j$, of the rotation group, \eg\ [\pref{Vilenkin, Miller, Hamermesh,
Bargmann}]. It is worth expanding on this in the following way.

A binary covariant, $K$, is an SL(2,C), and hence SU(2), invariant, \ie
(\cf\ (\peq{inv1})),
  $$
  K({\bf a};\xi,\eta)=K({\bf a}';\xi',\eta')\,.
  $$
Useful information follows from the infinitesimal expression of this
invariance.

The binary (spin--half) realisation of the generating angular momentum
operators is, (see Sharp, [\pref{Sharp}], Bargmann, [\pref{Bargmann}]),
  $$\eqalign{
  J_{+}&\to-\xi{\pa\over\pa\eta}\,,
  \quad J_-\to-\eta{\pa\over\pa\xi}\cr
  J_z&\to
  {1\over2}\bigg(\xi{\pa\over\pa\xi}-\eta{\pa\over\pa\eta}\bigg)\,,
  }
  \eql{binre}
  $$
acting on functions, $f(\xi,\eta)$. A commutation relation is
  $$
  [J_+,J_-]=2J_z\,.
  \eql{comm1}
  $$

For ease of comparison, I have reverted to the notation $J_{\pm}=J_x\pm
iJ_y$. These generators serve also for the binary invariant group, SL(2).

The total angular momentum (Casimir operator) is
  $$
  J^2={1\over4}\bigg(\xi{\pa\over\pa\xi}+\eta{\pa\over\pa\eta}\bigg)
  \bigg(\xi{\pa\over\pa\xi}+\eta{\pa\over\pa\eta}+2\bigg)\,,
  \eql{btot}
  $$
and the simultaneous eigenfunctions, $f_m^j$, of $J_z$ and $J^2$ are
easily determined by a standard procedure. From (\peq{btot}), they must
be homogeneous of degree $2j$, where $j(j+1)$ is the eigenvalue of $J^2$,
with $j$ a half--integer. The eigenvalue of $J_z$ equals $m$, with $-j\le
m\le j$. As usual, the $f_m^j$ can be determined, using the lowering
$J_-$, from the `highest weight' function $f_j^j$, which itself satisfies
$J_+\,f_j^j=0$ and so is fixed to be proportional to the power, $\xi^j$.
Then, lowering by induction, yields, see (\peq{nspin}),
  $$
  f^j_m=\xi^{(j)}_m\,,
  $$
and we see that the $\xi^{(j)}_m$ are function space variants of the
quantum vectors, $\ket {jm}$.

The matrix elements of $J_{\pm}$ in this basis are the usual ones. I do
not give them because they are contained in the expressions for the
transformations induced by (\peq{binre}) on the coefficients of the
quantic, (\peq{quantic}). These already occur in early invariant theory
in the guise of the Cayley--Sylvester--Aronhold operators,  $\Om$ and
${\it O}$.

Acting on a binary covariant, $K$, of the ground form, (\peq{quantic}),
(\peq{coeffc}), the operators $J_{\pm}$ are equivalent to the
annihilators, $\Om$ and ${\it O}$, \ie,
  $$\eqalign{
  \Om'K&\equiv\bigg(\Om-\eta{\pa\over\pa\xi}\bigg)K=0\cr
  {\it O}'K&\equiv\bigg({\it O}-\xi{\pa\over\pa\eta}\bigg)K=0
  }
  \eql{ann}
  $$
with, in classic notation,
  $$
  \Om\equiv \sum_{k=0}^{n-1}(k+1)\,a_k{\pa\over\pa a_{k+1}}\,,\quad
   {\it O}\equiv \sum_{k=0}^{n-1}(n-k)\,a_{k+1}{\pa\over\pa a_{k}}\,.
   \eql{ann2}
  $$

This theorem was proved by invoking the invariance of the covariant under
the infinitesimal shears, $\xi\to\xi+\la\eta$, $\eta\to\eta$ and
$\xi\to\xi$, $\eta\to\la\xi+\eta$, generated by $J_{-}$ and $J_{+}$,
respectively. (See \eg, Salmon, [\pref{Salmon}] \S148. Dickson,
[\pref{Dickson, Dickson2}], gives a succinct treatment and Elliott,
[\pref{Elliott}] Chaps.VI, VII a more lengthy one.) The operator $J_z$
generates the scalings that maintain $\xi\eta$. Olver, in his nice,
modern reworking of classical invariant theory, [\pref{Olver}], discusses
the Lie algebra aspects of $\Om$ and ${\it O}$.

Transcribed into SU(2) notation via (\peq{coeffs2}), equation
(\peq{ann2}) reads, ($n=2j$),
  $$\eqalign{
  \Om&=\sum_{m=1-j}^j\big((j-m+1)(j+m)\big)^{1/2}\,\phi^m{\pa\over\pa
  \phi^{m-1}}\cr
  {\it O}&=\sum_{m=1-j}^j\big((j-m+1)(j+m)\big)^{1/2}\,\phi^{m-1}{\pa\over\pa
  \phi^{m}}\cr
  &=-\sum_{m=-j}^{j-1}\big((j+m+1)(j-m)\big)^{1/2}\,\phi_{m+1}{\pa\over\pa
  \phi_{m}}\,,
  }
  \eql{omega}
  $$
in which the angular momentum matrix elements can be recognised using the
relation
  $$
   X_a=\varphi^i\big(G_a\big)_{i}^{\,\,j}{\pa\over\pa\varphi^j}
   \eql{greln}
  $$
between the Lie derivative operators, $X_a$, and the matrix generators,
$G_a$, in the (matrix) representation in the carrier space of which
$\varphi^i$ is a vector. Thus
  $$
  \varphi^j\to\,\,
  '\!\varphi^j=e^{\,q^aX_a}\,\varphi^j=\varphi^i
  \big[e^{q^aG_a}\big]_i^{\,\,j}\,.
  $$

The general structure of the SU(2) covariant, $K$, of degree $g$ and
order $\varpi=2s$ is a Clebsch--Gordan coupling of $g$ spin--$j$ objects,
$\phi$, to a spin-$s$ quantity (the set of coefficients) which is then
coupled to a spin-$s$ null spinor (the variables) to a zero spin
resultant. The invariance conditions (\peq{ann}) are then equivalent to
the standard invariance of the 3--$j$ symbols under SU(2).

The operators $\Om$ and ${\it O}$ perform important functions in the
construction of concomitants and thence in the proof of the Hilbert
finite basis theorem. The algebra of $\Om$ and ${\it O}$, developed for
this purpose, can be translated into angular momentum terms using
(\peq{omega}) and (\peq{greln}).

Corresponding to $J_z$, (\peq{comm1}), there is the commutator
(`alternant')
  $$
{1\over2}\, [{\it O},\Om]={1\over2}\sum_{k=0}^n(2k-n)a_k{\pa\over\pa
 a_k}=-\sum_{m=-j}^j m \,\phi^m{\pa\over\pa\phi^m}
 \eql{comm}
  $$
originally constructed by Cayley, [\pref{Cayley3}]. Therefore, in total,
$\Om'$, ${\it O}'$ and $\half[{\it O}',\Om']$ correspond to ${\bf
J}+{1\over2}\bsi$.\mgn{IS THIS TRUE?}

I reckon {\it weight} following Cayley's first scheme, [\pref{Cayley3}],
in which $a_r$ has weight $r-n/2$, and $\xi$ and $\eta$ have weights
$\pm1/2$. This is more in keeping with the group representation value
than the one usual in classical texts, being the eigenvalue of $J_z$, and
has the advantage that the commutator $\half[{\it O}',\Om']$ is the
weight, or scaling, operator since $\phi^m$ has weight $-m$.

A basic result, essentially due to Hilbert, is that the quantity,
  $$
  K({\bf a};\xi,\eta)=\bigg(1-{{\it O}'\Om'\over 1.2}
  +{{{\it O}'}^2{\Om'}^2\over 1.2^2.3}
  -{{{\it O}'}^3{\Om'}^3\over
  1.2^2.3^2.4}+\ldots\bigg)F({\bf a};\xi,\eta)\,,
  \eql{hproj}
  $$
where $F$ is a rational, integral, isobaric \footnote{ Isobaric means
that each separate term of an expression has the same weight. Olver
defines it to mean invariance under the scaling subgroup, which is more
restrictive.} function of the $a_r$ and $\xi,\eta$ of {\it zero} total
weight, is a covariant of the ground form (\peq{coeffc}), or
zero.\footnote { See Elliott, [\pref{Elliott}], \S182.}

The operator, $\caH_+$, in (\peq{hproj}), and its partner,
  $$
  \caH_-\equiv\bigg(1-{\Om'{\it O}'\over 1.2}+{{\Om'}^2{{\it O}'}^2\over 1.2^2.3}
  -{{\Om'}^3{{\it O}'}^3\over
  1.2^2.3^2.4}+\ldots\bigg)\,,
  \eql{hproj2}
  $$
take the forms of general projection operators constructed from raising
and lowering operators so that one could refer to $K$ as the {\it
covariant projection} of $F$. It is a theorem that every covariant can be
obtained in this way.

The operator equivalent of this projection, customarily couched in
angular momentum terms, is that the terminating polynomial,
  $$
  \sum_{r=0}^\infty{(-1)^r\over r!(1+r)!}\,J_-^r\,J_+^r
  \eql{low1}
  $$
is equivalent, using (\peq{comm1}), to
  $$
  \sum_{r=0}^\infty{(-1)^r\over r!(1+r)!}\,J_+^r\,J_-^r
  \eql{low2}
  $$
when acting on an eigenstate of $J_z$ with zero eigenvalue, $m=0$, and
projects onto the {\it identity}, or trivial, representation, \ie to a
scalar. This last statement accords with the scalar quality of the
covariant, $K$. The $m=0$ condition is just the zero weight one on $F$.

The operator, (\peq{low1}), is a special case of a projector derived by
L\"owdin, [\pref{Lowdin}], equn.(32) and rederived by Shapiro,
[\pref{Shapiro}]. This more general operator is equivalent, for quantics,
to ( $(\al)_n$ is the Pochhammer symbol),
  $$
  \bigg(1-{{\it O}\Om\over 1!(2m+2)_1}
  +{{{\it O}}^2\Om^2\over 2!(2m+3)_2}
  -{{{\it O}}^3\Om^3\over
  3!(2m+4)_3}+\ldots\bigg)\,,
  \eql{hproj3}
  $$
 where, for simplicity, I consider
actions on functions of just the coefficients, \eg\ on homogeneous,
isobaric functions \ie\ {\it gradients}, $G({\bf a})$, of {\it excess}
$2m$, a number which determines the scaling of $G$ via
  $$
  {1\over2}\,[{\it O},\Om]\,G=mG\,.
  $$
For these terms, consult Elliott, [\pref{Elliott}]. \footnote{
Elaboration of this will form part of a further communication.}

The sufficiency of dealing with just the projection onto the identity has
been exploited by Shapiro, [\pref{Shapiro2}], for arbitrary groups. The
basic notion is one commonly used in selection rule calculations that
computing how much of a given irrep there is in, say, a product of {\it
two} irreps, is equivalent to coupling the {\it three} irreps to the
trivial one. This corresponds to a simple group average, \eg\
[\pref{Hamermesh}]. It is the same as the choice one has in invariant
theory of dealing either with the form as a whole (a scalar) or with just
its coefficients (a carrier space vector).

As an example of a higher group, Noz and Shapiro, [\pref{NandS}], give
the identity projection for SU(3) as a polynomial in the generators. In
the invariant theory setting this involves an extension to the ternary
domain, geometrically more interesting, but more complicated. The most
systematic and elegant treatment of projection operators is given by
Story, [\pref{Story}], and a summary in [\pref{Story2}]. As an example,
the group GL(3) is generated by 6 shears (determinant one), and 3
expansions. In Elliott, [\pref{Elliott}], Chap.XVI, the shears are the
annihilators, denoted $\Om'_{ij}$ with $i,j=x,y,z;\, i\ne j$, and the
expansions are the non--zero commutators, $H'_1,H'_2,H'_3$ with $\sum
H'_i=0$. Elliott effectively works out the structure constants of
$\gs\gl(3)$ in the Cartan basis, the $H_i'$ being the Cartan subalgebra
in precisely the modern notation, \eg\ Racah, [\pref{Racah}], equn.(72).
For example, in terms of the Gell--Mann matrices $\la_3\sim H'_3$ and
$\la_8\sim H'_1-H'_2$ in the fundamental, quark representation,
$(x,y,z)$. $H'_1, H'_2$ and $H'_3$ generate scalings of the pairs
$(y,z)$, $(z,x)$ and $(x,y)$ respectively, corresponding to the three
SU(2) `binary' subgroups which give $V$--spin, $U$--spin and $I$--spin.
These scalings leave the products, $yz$, $zx$ and $xy$ invariant, each
pair as in the binary case, leading onto the definition of weights used
by Story.\footnote{ For those who wish to quickly refresh their knowledge
on the Lie algebra of GL(n)$\supset$ SL(n), I suggest Racah,
[\pref{Racah}], and Gourdin, [\pref{Gourdin}], esp. Chap.3.}

The projection operators are assembled out of the $\Om'_{ij}$, which are
the raising and lowering operators, usually denoted by $E_\al$ in Lie
algebra theory. A comparison of the projectors derived by Story,
[\pref{Story2}], equns.(8), (9) and Elliott, with those of Noz and
Shapiro, [\pref{NandS}], equns.(2.4), (2.17), shows agreement.\footnote{
This will be considered elsewhere.} See also Asherova and Smirnov,
[\pref{AandS}]. All methods involve chains of subgroups \eg\
SU(3)$\supset$SU(2).

There is a close connection between the construction of Lie group/algebra
representations and that of covariants from semi-covariants,\footnote{
These are the analogues of the extreme weight states, $\ket{jj}$ and
$\ket{j,-j}$.} in particular from one, given, term of the covariant (its
`source') by repeated action of $\Om'$ or ${\it O}'$, as in Roberts'
theorem in the binary domain. (I recommend [\pref{Dickson}],
[\pref{Elliott}] and [\pref{Hilbert2}]). The various matrix elements will
emerge as in (\peq{omega}) for $\gs\gl(2)$. The projection operator
method for simple Lie groups is given in more detail by A\v{s}erova {\it
et al}, [\pref{AST}].

\section{\bf 4. Polars and angular momentum addition.}

The polar operators ${\bf l\,.}\nabla_{\bf m}$, (\peq{polop}), and their
obvious extensions to higher domains, play important roles in invariant
theory.

The introduction of $u$ and $v$ is not necessary in the binary domain and
so is not usually made in the classical works. One can easily work
directly with $(-\eta,\xi)$. Line coordinates are needed however for
ternary and higher forms and their introduction in the binary case allows
a more unified treatment. \cf\ Todd, [\pref{Todd}],\S \S1.6, 2.12. For
example I now bring in the famous `Omega process' used in the composition
of invariants, and covariants, to give others. To be a little more
systematic, I denote point variables by $x_i$ ($i=1,\ldots,p$) and the
dual line coordinates by $u^i$. A $p$--ary $n$--ic is written
symbolically as,
  $$\eqalign{
   f:\quad a_x^n&=({\bf a.x})^n=(a^ix_i)^n=(a^1x_1+a^2x_2+\ldots+a^px_p)^n\cr
   F:\quad u_\al^n&=({\bf
u}.\bal)^n=(u^i\al_i)^n=(u^1\al_1+u^2\al_2+\ldots+u^p\al_p)^n
   }
  $$
in point and line coordinates respectively. I am most concerned with
$p=2$ and $p=3$.

The geometrical interpretation of forms becomes more involved the higher
$p$ is. One interpretation\footnote{ Not the only possibility.} follows
on taking the $x_i$ as homogeneous point coordinates in a
$(p-1)$--dimensional space and the vanishing of a form yields a
co--dimension 1 variety (a hypersurface) in this space. In the binary
case this is a range of points, in the ternary case, a curve and
quaternary, a surface. This interpretation provides a useful language and
visualisation, at least for low $p$.

Consider two distinct $p$--forms  of generally different orders,
$f=a_x^n\equiv u_\al^n=F$ and $g=b_x^m\equiv u_\be^m=G$, where I have
indicated both the point and line forms.

Following Clifford, [\pref{Clifford}], define the {\it generalised polar}
of $G$ in respect of $f$,
  $$\eqalign{
  h&={(n-m)!\over m!\,n!}\,\big(\nabla_u\,.\nabla_x\big)^m\,u_\be^m\,a_x^n
  \equiv{(n-m)!\over m!\,n!}\,\Om^m \,u_\be^m\,a_x^n \,\cr
  &={(n-m)!\over n!}\,\big(\bbe.\nabla_x\big)^m\,a_x^n\,\cr
   &=({\bf a\,.\,}\bbe)^m\,a_x^{n-m}= a_\bbe^m\,a_x^{n-m}
   \,,\quad n\ge m\,.
   }
   \eql{cgp1}
  $$
The first line is in Grace and Young (for the ternary case),
[\pref{GandY}], \S241, the second is Clifford's definition (also in the
ternary case). The last line is the symbolic expression. In the ternary
case, $h$ gives the {\it polar locus}, of order $(n-m)$, of $G$ in
respect of $f$.

$h$ is zero if $n<m$. However in this case, we can reverse the roles of
$f$ and $G$ and compute the generalised polar of $f$ in respect of $G$,
  $$\eqalign{
  h'&={(n-m)!\over m!\,n!}\big(\nabla_x\,.\nabla_u\big)^n\,u_\be^m\,a_x^n
  \equiv{(n-m)!\over m!\,n!}\,\Om^n \,u_\be^m\,a_x^n \,\cr
  &={(n-m)!\over m!}\,\big({\bf a}\,.\nabla_u\big)^n\,u_\be^m\,\cr
   &=({\bf a\,.\,}\bbe)^n\,u_\be^{m-n}=
   a_\be^n\,u_\be^{m-n}\,,\quad m\ge
   n\,,
   }
  \eql{cgp2}
  $$
which represents the {\it polar envelope}, of class $(m-n)$, of $f$ in
respect of $G$.\footnote{ A class curve is often called an envelope or a
tangential curve and an order curve is a `locus' A `curve' can usually be
expressed in either of these dual ways. The best known example is the
conic, which is of order 2, or equivalently, class 2. Simple
descriptions, in English, can be found in school textbooks such as
Robson, [\pref{Robson}], Milne, [\pref{Milnew}], Sommerville,
[\pref{Sommerville}]. Possibly the most attractive is Askwith,
[\pref{Askwith}]. More advanced is Salmon, [\pref{Salmon}]. The rather
rare book by Scott, [\pref{Scott}], should be mentioned and the more
modern Todd, [\pref{Todd}], and Semple and Kneebone, [\pref{SandK}].}

For binary forms, the situation is summarised, explicitly in point
coordinates, by Sturm, [\pref{Sturm}], who also describes the geometry in
terms of ranges of points on the complex projective line in a standard
way. For this purpose, it is expressive to rewrite $h$, say, using the
factored form of $u^m_\be$, which is, reverting to the original binary
notation,
  $$
  u^m_\be=\prod_{k=1}^m(v_k\,u-u_kv)\,,
  $$
so that setting $v_k\to\xi_k$ and $u_k\to-\eta_k$ and also, according to
the definition (\peq{cgp1}), $u\to\pa_\xi$, $v\to\pa_\eta$,
  $$
  h={(n-m)!\over
  n!}\,\prod_{k=1}^m\bigg(\xi_k{\pa\over\pa\xi}+\eta_k{\pa\over\pa\eta}\bigg)f\,,
  \eql{mixp}
  $$
which shows $h$ as a {\it mixed} polar of $f$ with respect to the points
(roots) $(\xi_k,\eta_k)$.

Writing $h$ as the form $({\bf c\,.\,x})^{n-m}$, the coefficients, $c_r$,
can be determined in terms of $a_r$ and $b_r$. They are given in Sturm,
for example, [\pref{Sturm}], and would allow one to calculate the 3$j$
symbols $\threej{n/2}{m/2}{(n-m)/2}***$.\footnote{ As might be expected,
these $3j$ symbols have simple explicit forms, \eg\ Hamermesh,
[\pref{Hamermesh}], p.375 eqn.(9-123). They occur later. }

In the binary case, the quantity $\Om$ is the operator introduced by
Cayley and which, in point coordinates, is expressed as a
determinant\footnote{ In the ternary case the operator $\Om$, as defined
here, occurs on p.296 of [\pref{GandY}] as $Q$.}. Use of line coordinates
allows it to be written as a contraction, $\Om=\nabla_u\,.\,\nabla_x$
(see Todd, [\pref{Todd}], \S2.12.3). Two binary forms, $f$ and $g$, can
be combined into a third using $\Om$ to give their $r$th {\it
transvectant},
  $$
  (f,g)^r={(n-r)!(m-r)!\over n!\,m!}\,\Om^r(fG)\,.
  \eql{transv}
  $$
In particular the polars are the `end values',
  $$
  h=(f,g)^m\,,\quad h'=(f,g)^n\,,
  $$
and generally $(f,g)^r$ is a form of order $(m+n-2r)$ on making the
replacement $(u_1,u_2)\to(-x_2,x_1)$ (or a form of class $(m+n-2r)$ on
making the dual replacement $(x_1,x_2)\to(u_2,-u_1)$. The order (class)
runs from $(m+n)$ to $|m-n|$ and transvection corresponds to addition of
angular momenta via Clebsch--Gordan, or 3$j$, symbols. As a well known
example, the second transvectant of a form with itself is proportional to
its {\it Hessian}, $H$, (\peq{hessian}).

I can now introduce the important notion of {\it apolarity}, originally
developed in the binary case but later extended to higher
domains.\footnote{ The notion appears to be due to Reye. I will not
dilate on its general significance but just refer to Coolidge,
[\pref{Coolidge}] pp.368, 410, Grace and Young, [\pref{GandY}], Semple
and Kneebone, [\pref{SandK}], Todd, [\pref{Todd}].}

The straight analytical definition is that two forms, $f$ and $G$, are
said \footnote{ Whether we say apolar to $G$ or to its dual, $g$, is,
metrically, immaterial and I will be somewhat slack in making this
distinction.} to be {\it apolar} if their polar, $h$, defined in
(\peq{cgp1}), vanishes. I take $n\ge m$ for ease.

An important fact is that the form $f$ is apolar to any form that
contains $G$ as a factor and whose class does not exceed $n$. The proofs,
which are simple, are given in Grace and Young, [\pref{GandY}] pp.225,
304. In the binary case this result corresponds to elementary angular
momentum addition rules, as I now explain. $f$ and $g$ correspond,
respectively, to spins $j_1=n/2$ and $j_2=m/2$ as explained earlier. The
polar $h$ corresponds to spin $j_1-j_2=(n-m)/2$. If it vanishes, so that
$f$ and $G$ are apolar, one has the 3$j$ equivalent
  $$
  f^{m_1}g^{m_2}\Threej{j_1}{j_2}{j_1-j_2}{m_1}{m_2}*=0\,,
  \eql{fgap}
  $$
\ie\ $2(j_1-j_2)+1=n-m+1$ conditions.

The angular momentum analogue of the product of two forms, say $g\,\phi$,
can be found from the multiplication rule for null spinors,
(\peq{ncomb}). Thus,
  $$\eqalign{
  g\,\phi&=g^{m_2}\,\phi^{\,m_3}\,\xi^{(j_2)}_{m_2}\,\xi^{(j_3)}_{m_3}\cr
  &=(-1)^{2j_2}(2j_2+2j_3+1)^{1/2}g^{m_2}\,\phi^{\,m_3}
  \Threej {j_2}{j_3}{m_4}{m_2}{m_3}{j_2+j_3}\, \xi^{(j_2+j_3)}_{m_4}\cr
  &\equiv\psi^{m_4}\,\xi^{(j_2+j_3)}_{m_4}\cr
  &=\psi\,.
  }
  \eql{prodf}
  $$
The statement is that $\psi,\,=g\phi,$ is apolar to $f$ if $g$ is, \ie,
that
  $$\eqalign{
 0&=  f^{m_1}\psi^{m_4}\Threej{j_1}{j_2+j_3}{(j_1-j_2-j_3)}{m_1}{m_4}*\cr
  &\propto f^{m_1}g^{m_2}\,\phi^{\,m_3}
  \Threej {j_2}{j_3}{m_4}{m_2}{m_3}{j_2+j_3}
  \Threej{j_1}{j_2+j_3}{(j_1-j_2-j_3)}{m_1}{m_4}*
   }
  $$
is true if (\peq{fgap}) holds. To show this, the 3$j$ symbols are
recoupled, using 6$j$ symbols, so that $j_1$ is coupled to $j_2$ and
$j_3$ to $j_1-j_2-j_3$. Without writing out the full expression, these
couplings quickly show that the smallest intermediate angular momentum is
$j_1-j_2$ and the largest one is also $j_1-j_2$. This allows (\peq{fgap})
to come into play and the result follows.

Two apolar forms of the same order (class), $n$, are sometimes called
{\it conjugate}.\footnote{ The terminology seems to originate with
Rosanes, [\pref{Rosanes}]. Reye uses the term apolarity as do Grace and
Young, [\pref{GandY}]. The classical expression for the joint invariant
(or {\it lineo linear invariant}) can be seen in many places \eg\
Elliott, [\pref{Elliott}], \S49, Grace and Young, [\pref{GandY}] \S177,
Glenn, [\pref{Glenn}], equn (71), Coolidge, [\pref{Coolidge}] p.369,
Rosanes, [\pref{Rosanes}]. The relation between (\peq{indt1}) and
(\peq{indt2}) is on p.58 of Elliott. Following Sylvester, Elliott
considers polars as examples of {\it emanants}.} From (\peq{fgap}), this
means that the joint invariant,
  $$
     f^{m_1}\,g_{m_1}=({\bf a\,.\,}\bbe)^n\,,
  \eql{jinv}
  $$
vanishes.

A form, $\phi$, of order $n$, is conjugate to the $n$th power (\cf\
(\peq{spsp})) of one of its factors, essentially by definition of the
roots of $\phi=0$. These powers are null forms, on my previous
definition, and their coefficients are the components of what I termed
the {\it principal null $2j+1$--spinors} of $\phi$, [\pref{DowkandG}].
{\it In general} there are $n=2j$ of them and any linear combination is
obviously conjugate to $\phi$. Conversely, any form conjugate to $\phi$
can be written as a linear combination of the principal null forms of
$\phi$, essentially by a counting argument (Rosanes, [\pref{Rosanes}]).

Since odd order forms (half--odd--integer spin) are self--conjugate, they
can always be expanded as sums of $n$th powers.

An easy theorem, which has sometimes been used as the definition of
apolarity, is that $f$ is apolar to $g$ if, and only if, it is conjugate
to the product $g\phi$ for {\it all} factors $\phi$. Thus, conjugacy
reads, using (\peq{prodf}) for the product of forms,
  $$
 f^{m_1} g^{m_2}\phi^{m_3}\Threej{j_2+j_3}{j_2}{j_3}{m_1}{m_2}{m_3}=0\,.
  $$
If true for all $\phi$ then $\phi$ can be removed and, setting $j_1\equiv
j_2+j_3$, there results
  $$
 f^{m_1}g^{m_2}\Threej{j_1}{j_2}{j_1-j_2}{m_1}{m_2}{m_3}=0\,.
  $$
This proves necessity. Sufficiency has been shown just above.

A specific example, that combines these lemmas, follows from the
groupings of the equation,
  $$
  a^{m_1}\,b^{m_2}\,c^{m_3}\Threej {j} {j-\half}{\half}{m_1}{m_2}{m_3}=0\,,
  \eql{combl}
  $$
which say, in form language, that, if $a_x^n$ and $b_x^{n-1}\,c_x$ are
conjugate $n$--ics, then $(ac)\,a_x^{n-1}$ and $b_x^{n-1}$ are conjugate
$(n-1)$--ics. The particular $3j$ symbol here is the same as the
matrix--spinor $u_A(j)$ ($A=\pm1/2$) employed by Dirac in his higher spin
equations (see Corson, [\pref{Corson}]).

As another illustration, in terms of $3j$ symbols the necessary and
sufficient condition for there to be at least $e$ equal roots of the
binary quantic equation $\Phi_{2j}=0$, is that the apolarity,
  $$
  \Threej{m''}{j'}j {j-j'}{m'}m\,\phi^m\,\xi^{m'}_{(j')}=0\,,\quad
  e=2(j-j')+1\,,
  \eql{eroots1}
  $$
should have solutions for $\xi$ corresponding to the equal root. This
implies
  $$
  \Threej{m''}{j'-\half}j {j-j'+\half}
  {m'}m\,\phi^m\,\xi^{m'}_{(j'-\half)}=\mu\,\xi^{m''}_{(j-j'+\half)}\,.
  \eql{eroots2}
  $$
In particular, as a trivial check, when $e=2j$, (\peq{eroots2}) implies
$\phi\propto\xi_{(j)}$, a null spinor. At the other extreme, if $e=1$
(\ie general position), (\peq{eroots1}) is just the polynomial,
$\Phi_{2j}=0$, each of the $2j$ roots qualifying as an `equal' root.

In general, (\peq{eroots1}) is a set of $e$ simultaneous $(2j+1-e)$-ics
with at least one common root, the equal root. If there are two common
roots, there are two sets of $e$ equal roots of the original quantic, and
so on.

As intimated earlier, for calculational purposes there is no need to
evaluate the $3j$--symbols in (\peq{eroots1}) as they are coded in the
polar, the derivative form of which, (\peq{cgp1}) or (\peq{mixp}), shows
that the $e$ polynomials are just derivatives applied to the form
$\Phi_{2j}$.

Explicitly (\cf\ Sturm, [\pref{Sturm}]), putting in a bookkeeping null
spinor, $\al_{(j-j')}$,
  $$
  \al_{(j-j')}^{m''}\,\xi^{m'}_{(j')}\!\Threej{j-j'}{j'}j {m''}{m'}m
  \phi^m=(-1)^{2j'}\!{(2j+1)^{1/2}\over(2j-2j')!}
  \bigg(\be{\pa\over\pa\xi}-\al{\pa\over\pa\eta}\bigg)^{\!2(j-j')}
  \!\Phi_{2j}(\ol\phi;\xi,\eta)\,.
  \eql{eroots3}
  $$

I also remark that the purely 2--spinor standpoint, (Penrose,
[\pref{Penrose2}]), is equally as effective.

To summarise the correspondances so far between angular momentum and
binary invariant theory: Forms of order (or class) $n$ correspond to
spin--$(n/2)$ spinors. The polar of one form with respect to another
corresponds to the addition of the angular momenta to give the minimum
total value. The product of two forms does the same thing but to the
maximum value, while the transvectant interpolates and gives all allowed
values.\footnote{ The connection of invariant theory and angular momentum
theory is, of course, well known but does not seem to be specifically
used, to any degree. Any necessary calculations tend to be done again.
However, for interesting modern developments, with an
algebraic--geometric slant, see Abdesselam and Chipalkatti,
[\pref{AandC}], and references given there. }

For the ternary case things are not so complete or straightforward but
they are geometrically more interesting. I will return to this topic
later and now take up the theory of harmonic polynomials, for which it is
pertinant.
\section{\bf 5. Harmonic projection. Poles.}
Historically, the Laplace coefficient $P_n(\cos\ga)$ arose in the Taylor
expansion \footnote{ I do not bother with conditions or regions of
convergence.} of the `displaced' $1/r$ potential, or static Green
function,
  $$\eqalign{
  {1\over|{\bf r-r'}|}&\equiv\sum_{n=0}^\infty{(rr')^n\,P_n(\cos\ga)\over r^{2n+1}}
  =\sum_{n=0}^\infty{H_n({\bf r,r'})\over r^{2n+1}}\cr
  &=e^{-{\bf r'}.\nabla}\,{1\over r}=\sum_{n=0}^\infty {(-1)^n\over n!}\,
  ({\bf r'.}\nabla)^n\, {1\over r}
  }
  \eql{Laplace}
  $$
yielding
  $$
  H_L({\bf r,r'})={(-1)^L\over L!}\, r^{2L+1}({\bf r'.}\nabla)^L\, {1\over
  r}\,,
  \eql{biax1}
  $$
given by Thomson and Tait, [\pref{TandT}] p.157 equn.(54). Symmetry
produces the equivalent expression,
  $$
  H_L({\bf r,r'})={(-1)^L\over L!}\, r'^{2L+1}({\bf r.}\nabla')^L\, {1\over
  r'}\,.
   \eql{biax2}
  $$

A form which displays the symmetry can be found by remarking that, as we
know (\peq{cmn}), the left--hand side of (\peq{cmn}),
$(\goa_xx+\goa_yy+\goa_zz)^L$, is a spherical harmonic, of degree $L$.
This can be checked by direct differentiation, assuming the null
condition, $\goa\,.\,\goa=0$. Hence, in particular,
  $$
  ({\bf r.}\nabla')^L\,{1\over r'}
  \eql{harm1}
  $$
is a spherical harmonic in ${\bf r}$, in agreement with (\peq{biax2}) of
course. The next step involves an application to (\peq{harm1}) of the
useful differentiation theorem, due to Niven, [\pref{Niven}], see Hobson,
[\pref{Hobson}], and [\pref{BandL}] (6.169),
  $$
 Y_n({\bf r})= (-1)^n{ 2^nn!\over (2n)!}\, r^{2n+1}\,Y_n(\nabla)\,{1\over r}\,,
  \eql{difft1}
  $$
where $Y_n$ is any solid spherical harmonic. There results Niven's
interesting form,
  $$
  H_L({\bf r,r'})={2^L(rr')^{2L+1}\over(2L)!}\,(\nabla.\nabla')^L{1\over
  rr'}\,.
  \eql{niven1}
  $$
I note the similarity of the equivalent expressions, ({\peq{niven1}) and
(\peq{biax1}), with the polar forms in (\peq{cgp2}).

The expression of spherical harmonics as derivatives, normal {\it or}
fractional, of $1/r$ was taken as fundamental by Thomson and Tait,
[\pref{TandT}], and also was adopted and extended by Maxwell,
[\pref{Maxwell}].

In this connection, it is best to begin historically with what might be
called Gauss' harmonic expansion. Gauss proved  constructively, in a
brief `Nachlass', that a rational, integral polynomial function, of
degree $n$, $f_n({\bf r})$, can be represented by a finite sum of solid
spherical harmonics, $Y_m({\bf r})$, as,
  $$\eqalign{
   f_n&=Y_n+r^2\,Y_{n-2}+r^4\,Y_{n-4}+\ldots\cr
       &=\sum_{s=0}^{[n/2]}r^{2s}\,Y_{n-2s}\,.
  }
  \eql{Gauss}
  $$

By repeatedly hitting the left--hand side with the Laplacian until zero
is obtained and then solving the resulting equations from the bottom up,
all the harmonics, $Y_m$ can be found in terms of $f_n$.\footnote{ A
specific example is worked out by MacMillan, [\pref{MacMillan}], \S207.
See also Heine, [\pref{Heine}] vol.1, pp.324--325. } The expansion of a
given $f_n$ is, therefore, unique.

To find an explicit expression for the $Y_m$ in terms of $f_n$, which
would be desirable, one can start by obtaining the harmonic, $Y_n$, from
$f_n$. This can be found in the work by Clebsch, [\pref{Clebsch}]. The
relevant theorem, in the way he states it, can be motivated, in our
terms, by writing (\peq{Gauss}] as,
  $$\eqalign{
   f_n&=Y_n+r^2\big(Y_{n-2}+r^2\,Y_{n-4}+\ldots\big)\cr
  &=Y_n+r^2\,f_{n-2}\,,
  }
  \eql{Gauss2}
  $$
where the first line is the iteration of the second. This formula
expresses the standard decomposition of (the space of) homogeneous
polynomials and is usually derived first (\eg\ Vilenkin,
[\pref{Vilenkin}], Berger {\it et al}, [\pref{BGM}]) with the expansion
(\peq{Gauss}) as a consequence).

One refers to $Y_n$ as the {\it harmonic projection} of $f_n$,
$Y_n=H(f_n)$, and Gauss' expansion can be construed as a (finite) series
of harmonic projections,
  $$
   f_n=H(f_n)+r^2\,H(f_{n-2})+ r^4\,H(f_{n-4})+\ldots
   \eql{harms}
  $$

Clebsch, [\pref{Clebsch}] eqn.3, gives the harmonic projection, although
he does not use this terminology, in the form,\footnote{ A more
sophisticated treatment is given in Vilenkin, [\pref{Vilenkin}] Chap.XI.
Also the quantities $r^2$ and $\De$ are reciprocal, in the sense of
Salmon, and the projection operator takes a typical form. Compare with
the Hilbert projectors, (\peq{hproj}) (\peq{hproj2}) (\peq{hproj3}).
$r^2$ is a raising, and $\De$ a lowering, operator, \cf\ Elliott,
[\pref{Elliott2}]. The Lie algebra aspect of spherical harmonics is
outlined by Howe, [\pref{Howe}], as an example of a bigger scheme. His
other examples are also instructive.}
  $$
 Y_m= H(f_m)\equiv \bigg[1-{r^2\,\De\over2\,(2m-1)}+{r^4\,\De^2
  \over2.4\,(2m-1)(2m-3)}-\ldots\bigg]\,f_m\,.
  \eql{harmp2}
  $$

Exhibiting all the projections, $H(f_m)$, in terms of the starting out
function, $f_n$, will complete the evaluation. To do this, I follow the
Gauss procedure, by writing out the series again,
  $$
  f_n=Y_n+r^2\,Y_{n-2}+\ldots+r^{2s}\,Y_{n-2s}+r^{2s+2}\,Y_{n-2s-2}+\ldots\,,
  $$
and then acting on it with $\De^s$, which kills the first $s$ terms. This
leaves
  $$
  \De^s\,f_n=A_s(n)\,Y_{n-2s}+B\,r^2Y_{n-2s-2}+\ldots\,
  $$
where $A_s(n)$ is the constant,
  $$
  A_s(n)=2.4.\ldots2s \,(2n-2s+1)(2n-2s-1)\ldots(2n-4s+3)
  \eql{coeff2}
  $$
and $B$ is a constant that I do need because the harmonic (first) part of
the equation gives,
  $$
  Y_{n-2s}\equiv H(f_{n-2s})={1\over A_s(n)}\,H(\De^s\,f_n)\,.
  \eql{harmp}
  $$

In this way, all the terms in (\peq{harms}) can be considered calculated.
Equation (\peq{harmp}) with (\peq{coeff2}), was obtained by Prasad,
[\pref{Prasad}], and later by Dougall, [\pref{Dougall}] \footnote{ There
is an ellipsis of an ellipsis in the expression for $C_{2p}$ in Hobson
[\pref{Hobson}], \S96.}.

Prasad goes on to express the expansion in a compact way using the more
general differentiation theorem of Hobson, [\pref{Hobson1}],
[\pref{Hobson}], \S80,
  $$
 H(f_n)=(-1)^n {2^n\,n!\over(2n)!}\, r^{2n+1}\, f_n(\nabla){1\over r}\,,
  \eql{difft2}
  $$
which is an extension of Clebsch's result, (\peq{harmp2}). (The more
specific result (\peq{difft1}) follows from (\peq{difft2}).) Easy
substitution yields the final expansion,
  $$
  f_n({\bf
  r})=(-1)^n\sum_{s=0}^{[n/2]}{(2n-4s+1)\over A_s(n)}\,
 r^{2n-2s+1}\big[\De_\rho^s\,f_n({\brho})\big]{1\over r}\,,
  $$
where $\brho$ is the vector of gradients,
$\brho=(\al,\be,\ga)=(\pa_x,\pa_y,\pa_z)$, and
  $$
  \De_\rho={\pa^2\over\pa\al^2}+{\pa^2\over\pa\be^2}+{\pa^2\over\pa\ga^2}\,.
  $$

As noted by Clebsch, specific choices of $f_n$, produce known harmonics.
For example
 $$
  f_n({\bf r})=C\,({\bf r.r'})^n\,,\quad C=2^{-2n}{(2n)!\over n!n!}=
  {1.3.\ldots(2n-1)\over n!}
  \eql{fp}
  $$
yields the usual Laplace coefficients, $P_n(\cos\ga)$, as follows
immediately from (\peq{difft2}) and (\peq{biax1}). Hobson,
[\pref{Hobson}], uses (\peq{difft2}) to develope the properties of the
zonal and tesseral spherical harmonics following Maxwell, and says
everything that needs saying.

Now $({\bf r.r'})^n$ is a rather special $n$th order polynomial (in ${\bf
r}$) which suggests a natural extension to the {\it general} product of
$n$ factors,\footnote{ Usually the vectors ${\bf p}_{(i)}$ are taken to
be unit ones.}
  $$
  f_n({\bf r})=C\,\prod_{i=1}^n\,\big({\bf p}_{(i)}\,.\,{\bf r}\big)\,,
  \quad {\bf p}_{(i)}=(x_{(i)},y_{(i)},z_{(i)})
  \eql{prod3}
  $$
whose harmonic projection, with $C$ as in (\peq{fp}), gives the {\it
solid} spherical harmonic
  $$\eqalign{
  Y_n({\bf r},\widehat{\bf p}_{(i)})&={(-1)^n\over n!}\,r^{2n+1}\,
  \prod_{i=1}^n\,\big({\bf p}_{(i)}\,.\,{\bf \nabla}\big)\,{1\over r}\cr
  &=
   {(-1)^n\over n!}\,r^{2n+1}\,
  \prod_{i=1}^n\,\nabla_{{\bf p}_{(i)}}\,{1\over r}\,,
  }
  \eql{polharm}
  $$
where $\nabla_{{\bf p}_{(i)}}$ is a directed derivative.

Maxwell refers to the points where the rays, determined by the vectors
${\bf p}_{(i)}$, intersect the unit 2-sphere as the {\it poles} of the
harmonic, $Y_n$, and considers the harmonic to depend on them. I have
indicated this by the dependence on the unit vectors $\widehat{\bf
p}_{(i)}$. As noted by Maxwell, Gauss, in the same Nachlass referred to
earlier, sketchily outlines a geometrical meaning of `Sphere functions'
referring to the poles as `bestimmte Punkte'.

Conversely, by a non--rigorous counting method, Maxwell shows that {\it
any} rational integral harmonic, $Y_n$ of order $n$ can be represented as
a multipole, (\peq{polharm}). This is an important point and is the
content of Sylvester's theorem.
\section{\bf 6. Clebsch--Sylvester theorem.}
Proceeding a little more carefully, \cf\ [\pref{Hobson}],
[\pref{Hobson1}], [\pref{CoandH}], [\pref{Lense}] \S 51, the question is,
{\it given} a specific harmonic, $Y_n$, can one find a general
homogeneous polynomial, $f_n$, of product form
  $$
  f_n({\bf r})=C\,\prod_{i=1}^n ({\bf p}_{(i)}\,.\,{\bf r})\,,
  \eql{prod1}
  $$
of which $Y_n$ is the harmonic projection?

The relation between $Y_n$ and $f_n$ is the basic Gauss expansion,
(\peq{Gauss2}). We see that this does not determine $f_n$ uniquely
because of the the final term, which can be chosen arbitrarily, given
just $Y_n$. It is this `gauge' freedom that can be used to work $f_n$
into the product form, (\peq{prod1}). Saying it again, we are looking to
write the harmonic $Y_n$ is the form,
  $$
  Y_n({\bf r})=C\,\prod_{i=1}^n ({\bf p}_{(i)}\,.\,{\bf r})+r^2 G_{n-2}({\bf
  r})\,,
  \eql{prod2}
  $$
by appropriate choice of the general homogeneous polynomial of order
$(n-2)$, $G_{n-2}$. This is a purely algebraic problem. first raised, and
solved, by Clebsch and, later clearly, but more cursorily, by Sylvester.
The question is independent of the notion of poles.

Interpreting $(x,y,z)$ as projective (homogeneous) coordinates in the
plane, the terms in (\peq{prod2}), correspond to lines $h_i$, ${\bf
p}_{(i)}\,.\,{\bf r}=0$, a curve, $C_n$, of order $n$, $Y_n({\bf r})=0$,
a curve, $C_{n-2}$, of order $(n-2)$, $G_{n-2}=0,$ and an absolute null
conic, $C_2$, $r^2={\bf r\,.\,r}=x^2+y^2+z^2=0$. Geometrically
(\peq{prod2}) says that the $n$ lines $h_i$ meet the curve $C_n$ in the
$2n$ points where the conic $C_2$ cuts that curve. These points occur in
$n$ complex conjugate pairs and it is only lines joining such pairs that
are {\it real} lines, and so can be identified with $h_i$. There is
therefore only one way of doing this, so that the decomposition into a
real product is unique. This is commonly referred to as {\it Sylvester's
theorem} but there is a very clear and explicit statement of it in
Clebsch, [\pref{Clebsch}] p.350, and so I suggest it be called the
Clebsch--Sylvester theorem. The problem may be that Clebsch uses a rather
complicated elimination procedure.

A further piece of old fashioned geometric terminology is to say that the
line, $h$, is the polar, with respect to the null conic, of the point
(the pole) whose homogeneous coordinates are the line coordinates
(direction cosines) of $h$.\footnote{ There seems to be a curious
congruence of terms as the word pole is used both in its projective
geometric sense and in its physical sense, although both derive from the
same root meaning `axis'.}

The determination of the intersections, \ie of the poles, by elimination,
is a separate algebraic matter best approached by first using the Cartan
map to parametrise, this time, the variables, $(x,y,z)$, rather than the
coefficients, of the ternary $n$--ic, $Y_n$,
  $$
  Y_n(x,y,z)=\sum {n!\over p!q!r!}\,a_{pqr}\,x^py^qz^r;\quad p+q+r=n\,,
  \eql{tnic}
  $$
so that the null conic, $C_2$, is automatically satisfied.

In my previous notation I denoted the {\it null} ${\bf r}$ by $\gob$,
with the Cartan map, or {\it uniformisation} in algebraic geometry,

$$
   \gob_1= - {\gob_x-i\gob_y\over i\sqrt2}=\xi^2\,\,,\quad \gob_{-1}=
    {\gob_x+i\gob_y\over i\sqrt2}=\eta^2\,
    \,,\quad\,\gob_0=-i\gob_z=\sqrt2\,\xi\eta\,,
   \eql{cm2}
    $$
which differs, only by conventions, from existing choices,
[\pref{CoandH}], [\pref{Lense}].

Before pursuing the required elimination, I wish to make some comments
related to (\peq{cm2}).  Substitution of (\peq{cm2}) into the ternary
$n$--ic, (\peq{tnic}), yields a {\it binary} $2n$--ic,
  $$
  Y_n(\gob_x,\gob_y,\gob_z)=Y_{2n}(\xi,\eta)=
  \sum{(2n)!\over p!\,q!}\,a_{pq}\,\xi^p\,\eta^q;\quad
 p+q=2n\,.
  \eql{2ys}
  $$

Equation (\peq{nn}) can be considered as a formal expression of this
equivalence \footnote{ The equivalence between the invariant theory of a
binary quantic and an orthogonal ternary quantic, which is what we really
have here, has been utilised by Littlewood [\pref{Littlewood}]. He makes
no mention of the work of Cartan on spinors nor of the relevant
considerations of Burnside and Salmon.} through the equality of the
symbolic forms of the two types of quantics. In fact it is helpful, in
the present situation, to draw attention to the similarity between the
true power $({\bf r'\,.\,r})^n$ of (\peq{fp}), whose harmonic projection
gives $P_n$, and the symbolic form $(\goa\,.\,{\bf r})^n$ of a general
ternary quantic.\footnote{ See Problem 9, [\pref{Hobson}], p.176. \cf\
[\pref{Clebsch}].} Then, still symbolically, the harmonic projection
yields
  $$\eqalign{
  Y_n({\bf r},\goa)&={(-1)^n\over n!}\,r^{2n+1}\,
  (\goa\,.\,{\bf \nabla})^n\,{1\over r}\cr
  &=
   {(-1)^n\over n!}\,r^{2n+1}\,
  \,\nabla_{\goa}^n\,{1\over r}\,.
  }
  \eql{polharm3}
  $$
\section{\bf 7. Computation of the poles.}
The elimination, to find the factors, \ie the poles, is reduced to
solving the $2n$--ic,
   $$
   Y_{2n}(\xi,\eta)=0\,,
   \eql{poly3}
   $$
for the ratio $t=\xi/\eta$. From reality, for every root, $t$, there is
another equal to $-1/t^*$. This means that the solutions for the
corresponding vector $\gob$ (projectively a point) are complex
conjugates, as mentioned above. The line, $h$, joining two such points
is,
  $$
  \left|\matrix{x&y&z\cr
                \gob_x&\gob_y&\gob_z\cr
             \gob^*_x&\gob^*_y&\gob^*_z\cr}\right|=0
  \eql{lines}
  $$
or
  $$
   h:\,\quad{\bf r}.(\gob\times\gob^*)=0\,,
  $$
so that the real pole vectors, ${\bf p}_{(i)}$, in (\peq{prod2}) are,
  $$
   {\bf p}_{(i)}=\pm i\,\gob\times\gob^*\,,
  $$
for every pair of null solutions $\gob,\gob^*$. The direction cosines
(line coordinates) are
  $$
   \widehat{\bf p}_{(i)}=\pm i\,{\gob\times\gob^*\over\gob\,.\,\gob^*}\,.
  \eql{dirc}
  $$

The spherical harmonics, (\peq{polharm}) now look like,
   $$\eqalign{
  Y_n({\bf r},\{\gob\} )&=C\,r^{2n+1}\,
  \prod_\gob\,\big(\gob^*\,.\,(\gob\times i\,\nabla)\big)\,{1\over r}\cr
   &=C\,r^{2n+1}\,\prod_\gob
   \left|\matrix{i\pa_x&i\pa_y&i\pa_z\cr
                \gob_x&\gob_y&\gob_z\cr
             \gob^*_x&\gob^*_y&\gob^*_z\cr}\right|{1\over r}\,.
  }
  \eql{polharm4}
  $$

As an example take the real harmonic polynomial,
  $$
  H({\bf r})=-3x^4-3y^4-8z^4-6x^2y^2+24y^2z^2+24x^2z^2-60\sqrt2x^2yz+20\sqrt2y^3z
  \,.
  $$
Replace ${\bf r}$ by $\gob$ and then use the Cartan map to get the
octavic (without an 8th power),
  $$
  H(\xi,\eta)=20\xi\eta\big(2i\sqrt2(\xi^6+\eta^6)-7\eta^3\xi^3\big)=0\,,
  $$
whose roots are easily obtained. I give only one,
  $$
  \xi/\eta=-{1\over2}\sqrt{3\over2}+{i\over2\sqrt2}\,.
  $$
Substitution back into the Cartan map yields the null vector,
  $$
  \gob=\bigg(-{\sqrt6\over8}+{3\sqrt2\,i\over8},\,\,
  {5\sqrt2\over8}-{\sqrt6\,i\over8},\,\,
  -{1\over2}-{\sqrt3\,i\over2}\bigg)\,,
  $$
and evaluation of the direction cosines (\peq{dirc}) gives,
  $$
   \widehat{\bf
p}_{(1)}=\bigg(-\sqrt{2\over3},\,\,-{\sqrt2\over3},\,\,-{1\over3}\bigg)\,.
  $$

In fact, in this case, the poles are the corners of a regular
tetrahedron, centred on the origin with one corner on the $z$--axis, one
lying in the $xy$--plane and one edge parallel to the
$x$--axis.\footnote{ The converse constructive calculation of $H({\bf
r})$ is given in MacMillan, [\pref{MacMillan}], p.397.}

The computation of the null vectors can be bypassed by going back
essentially to the factorisations and the basic equality (\peq{2ys})
which is,
  $$
  \prod_{i=1}^n(\gob\,.\,{\bf p}_{(i)})=\eta^{2n}
  \prod_{i=1}^{n}\big(t-t_i\big)\bigg(t+{1\over
  t_i^*}\bigg)\,.
  \eql{fact}
  $$
On rewriting both sides, the left one best in spherical coordinates,
$\gob_1=\xi^2=\eta^2\,t^2$, $\gob_{-1}=\eta^2$,
$\gob_0=\sqrt2\xi\eta=\sqrt2\,\eta^2\,t$, one finds,
  $$\eqalign{
  \prod_{i=1}^n\big(\gob_m\,{\bf p}_{(i)}^m\big)
  &=\eta^{2n}\prod_{i=1}^n\big(t^2\,
  {\bf p}_{(i)}^{1}+\sqrt2\,t\,{\bf p}_{(i)}^{0}+\,{\bf p}_{(i)}^{-1}\big)\cr
  &=\eta^{2n}\prod_{i=1}^{n}\bigg(t^2+{1-|t|^2)\over|t|^2}
  \,t_i\,t-{t_i^2\over|t|^2}
  \bigg)\,.
}
  \eql{fact1}
  $$

From this identity the values of the spherical (contrastandard)
components of the pole vectors can be read off. When normalised, I find
  $$
  \widehat{\bf p}_{(t)}=-{\sqrt2i\over1+|t|^2}
  \bigg(t^*,\, {1\over\sqrt2}\,(1-|t|^2),\,\,
  t\,\bigg)\,,
  \eql{dirc3}
  $$
where I have labelled the pole by the root, $t$ (dropping the index
`$i$'). The transformation $t\to-1/t^*$ just reverses the direction of
the ray, giving nothing new.

For convenience, I give the Cartesian components as well,
  $$
  \widehat{\bf p}_{(t)}={2\over1+|t|^2}\bigg(\Real t,\,
-\Imag t,\, -{1\over2}\,(1-|t|^2)\bigg)\,.
  \eql{dirc2}
  $$

The use of the Cartan map to solve the elimination problem, leading to
the result (\peq{dirc2}), is probably the most efficient technique of
finding the poles. It occurs in Backus, [\pref{Backus}], and Baerheim,
[\pref{Baerheim}]. The articles of Katz and Weeks, [\pref{KandW}] and
Weeks, [\pref{Weeks1}] contain a handy independent reworking of the
basics and some relevant numerics.

After several steps of elimination, not directly involving the Cartan
map, Clebsch, [\pref{Clebsch}], also provides a means of finding the
poles by first factorising a polynomial constructed from the harmonic,
giving what I have denoted by $\gob$. The poles are then obtained from
(\peq{lines}).

More explicitly, in his notation, $\Om$ is a homogeneous harmonic
polynomial of order $n$, and Clebsch proves that the (finite) expression,
of order $2n$,
  $$
  \Upsilon\equiv\Om^2-{n\over n} {r^2\over 2!}\,\De\,\Om^2+
  {n(n+1)\over n(n-1)} {r^4\over 4!}\,\De^2\,\Om^2-\ldots \,,
  \eql{Clebsch2}
  $$
is always factorisable into $2n$ linear factors.\footnote{ There is a
misprint on p.349 in [\pref{Clebsch}], $n$ factors being stated.} These
are the $\gob$s.

As an example I choose the simple, $n=2$ harmonic, $\Om=xy+yz+zx$, and
find,
  $$\eqalign{
  \Upsilon&=\!x^4\!-\!2x^3y\!-\!2x^3z\!+\!3x^2y^2\!
  +\!3x^2z^2\!-\!2xy^3\!-\!2xz^3+\!y^4
  -2y^3z+3y^2z^2-2yz^3+z^4\cr
  &=\!(x^2+y^2+z^2-xy-yz-zx)^2\cr
  &=\!\bigg(x-{y+z\over2}+{\sqrt3 i|y-z|\over2}\bigg)^2
  \bigg(x-{y+z\over2}-{\sqrt3\, i|y-z|\over2}\bigg)^2\,,
  }
  \eql{Clebsch3}
  $$
giving the null vector (in Cartesians),
  $$
  \gob\sim\bigg(1,-{1-\sqrt3\,i\over2},-{1+\sqrt3\,i\over2}\bigg),
  $$
and its complex conjugate, twice. Computing the poles via (\peq{dirc})
gives ${1\over\sqrt3}(1,1,1)$, twice. For comparison, the alternative
method can be seen later, in connection with another matter.

In the general situation, the linear factors can be found by first
setting $z=1$ and then, in crudest fashion, substituting $y=mx+c$ into
the identity, $\Upsilon=0$ to give $\Upsilon(x)=0,\,\forall x$. In fact,
one needs only two terms, say $\Upsilon(0)=0$ and $\Upsilon'(0)=0$. The
former determines $c$ as the solution of an ordinary polynomial of degree
$2n$ and the latter fixes $m$ linearly in terms of $c$. This method is,
perhaps, less efficient than the one based on the binary form because of
the redundant information.

\section{\bf 8. Normal forms.}

In many cases, the harmonic polynomial will be presented as a combination
of the standard set of spherical harmonics, $C^L_M({\bf r})$. This is a
`normal form' of the harmonic, [\pref{Hobson}] \S87. Starting from this
combination allows a re--expression of the factorisation theorem.

For a specific order $L$, let the harmonic,
   $$
   \Phi_L({\bf r})=\phi_L^M\,C^L_M({\bf r})\,,
   \eql{phi1}
   $$
be the real data under consideration. (I called this $Y_n({\bf r})$
before.) Using
  $$
  \big(C^L_M({\bf r})\big)^*=C_L^M({\bf r})\,,
  \eql{ccsp}
  $$
reality implies the condition,
  $$
  {\phi^*}^M_L=(-1)^{L-M}\,\phi^{-M}_L=\phi^L_M\,,
  \eql{cond2}
  $$
familiar from mode decompositions into spherical harmonics in many
physical situations such as hydrodynamics. In the terminology of
[\pref{DowkandG}], the raising and lowering operator is a charge
conjugation operator.

Replacing ${\bf r}$ by a null vector, $\gob\in\oC^3$, the real harmonic,
(\peq{phi1}), turns into,
  $$
  \Phi_L(\gob)=\phi_L^M\,C^L_M(\gob)\,,
  \eql{phi2}
  $$
still with the reality condition (\peq{cond2}) on the data. Quite
generally, one has  $\Phi_L(\gob)^*= \Phi_L(\gob^*)$, which shows that if
$\gob$ is a solution, \ie if
  $$
   \Phi_L(\gob)=0\,,
  $$
then so is $\pm\gob^*$, as Clebsch and Sylvester said.\footnote{ For
clarity, $\gob^*$ here refers to the vector with complex conjugated
Cartesian components.}

It is a useful exercise to confirm this, since the replacement ${\bf
r}\to\gob$, modifies the reality properties, (\peq{ccsp}), of the
harmonics. One can use either $\gob$ or $(\xi,\eta)$ of (\peq{cm2}) to
express this. We also have the explicit formulae (\peq{nsph}) and
(\peq{nspin}) to check things out.

Taking the complex conjugate of $\Phi_L$ gives,
  $$\eqalign{
  \Phi_L(\gob)^*&={\phi^*}_L^M\,\big(C^L_{M}(\gob)\big)^*=
  (-1)^{L+M}\,\phi^{-M}_L\big(C^L_{M}(\gob)\big)^*\cr
  &=(-1)^{L-M}\,\phi^{M}_L\big(C^L_{-M}(\gob)\big)^*\cr
  &=\phi^{M}_LC^L_{M}(\gob^*)\cr
  &=\Phi_L(\gob^*)\,.
  }
  $$

$\Phi_L$ is an easily found binary \footnote{ It is of course also a
ternary form as mentioned earlier.} $2L$-ic in the variables $\xi$ and
$\eta$ (\cf\ equations (\peq{quantic}) and (\peq{2ys}) in a different
notation) whose roots correspond to the Maxwell poles.\footnote{ To
determine the poles, it is not necessary to convert this to a polynomial
in $(x,y,z)$ as in Weeks, [\pref{Weeks1}].} Although the algebra and the
properties of the polynomial have been given before, I rederive them in a
not really different way.

Looking at the polynomial equation (equivalent to (\peq{poly3})) as one
in $t=\xi/\eta$, the replacement $\gob\to\gob^*$ corresponds to charge
conjugation, $\xi\to-\eta^*,\,\,\eta\to\xi^*$, and so, if $t$ is a root,
then so is $-1/t^*$, as I said before. I have thus obtained the
factorisation displayed on the right--hand side of (\peq{fact}). To
determine the poles, this 2--spinor factorisation is simply rewritten in
terms of vectors, as on the left--hand side of (\peq{fact}).

The spinor $\psi^{\dag}=(-\eta^*,\xi^*)$ is the {\it spin conjugate}
spinor to $\psi=\big({\xi\atop\eta}\big)$ and one can refer to the
complex conjugate null 3--vector $\gob^*$ as the conjugate vector
$\gob^{\dag}=\gob^*$.\footnote{ A detailed discussion of the notion of
spinor and null vector in connection with rotations in 3--space can be
found in Kramers' nice book on quantum mechanics, [\pref{Kramers}], based
on his earlier papers, [\pref{Kramers1}]. The definitions mean that his
null vectors are $i$ times mine. He also usefully extends to the
relativistic case. His approach can be found more fully in Brinkman,
[\pref{Brinkman}].}

I remark that this analysis depends on the fact that the harmonic data is
real. I also remark that pairing other roots produces other, complex,
vector factors. The non-triviality of the result lies in the fact that
the factors are real (if $\Phi$ is). Any binary quantic can be factorised
and hence so can $\Phi(\gob)$, but with complex factors, in general. Some
further remarks on this can be found in [\pref{Lachieze}].

If ${\bf r}$ is reinstated, the factorisation statement, (\peq{prod2}),
is that the harmonic polynomial $\Phi_L({\bf r})$ can be decomposed as
  $$
  \Phi_L({\bf r})=C\prod_{i=1}^L({\bf r}\,.\,{\bf p}_{(i)})+r^2G_{L-2}({\bf
  r})\,.
  \eql{decomp}
  $$

Both the polynomial, $G_{L-2}$, and the constant, $C$, up to scaling of
the ${\bf p}_{(i)}$, are uniquely determined by $\Phi$.

The constant $C$ can be found by substituting a value for ${\bf r}$ that
lies on the conic and then $G$ follows by subtraction. It can also be
constructed from the poles, as will be shown later.

I also make the important point that the factors are real, if $f_n$ is.

For completeness, the constant $C$ can also be determined as follows,
(\eg\ [\pref{Lense}], \S 51, [\pref{CoandH}]). Consider the pencil of
curves,
  $$
   \Phi_L({\bf r})-\la\,\prod_{i=1}^L\big({\bf r}\,.\,{\bf
   p}_{(i)}\big)=0\,,
  \eql{bun}
  $$
for varying $\la$. Each member of this pencil goes through the $2L$
intersections (of $C_2$ and $C_L$). Select a point, not one of these
intersections, on the conic, $C_2$, and adjust $\la$ so that the
corresponding curve of the pencil goes through it, which is possible
because none of the lines ${\bf r\,.\,p}_{(i)}=0$ vanishes at this point.
Since this curve of the pencil now has $2L+1$ points in common with the
absolute conic, $C_2$, it must completely contain it, and so degenerate
into the conic and a curve of the $(n-2)$nd order, $C_{n-2}$. This
implies that the left--hand side of (\peq{bun}) must have ${\bf r}^2$ as
a factor, the sufficiency of which is obvious. (A proof of the necessity
was given by Ostrowski, [\pref{Ostrowski}], [\pref{CoandH}].) The so
determined $\la$ is the sought for constant, $C$, the value being
independent of the initial point selected.
\section{\bf 9. Apolarity and harmonic polynomials.}
This development can be given an older analytical--geometric terminology
and, to this end, I return to invariant theory. The relevant statement is
that if a homogeneous polynomial in $(x,y,z)$ (a ternary form) is
harmonic, it is apolar to the fundamental (null) conic, $C_2$, and
conversely. This is easily shown from the definition (\peq{cgp1}) which
says that the polar is obtained by replacing the line coordinates, ${\bf
u}$, by the derivative, $\nabla$, in the envelope equation of the conic,
and letting it act on the ternary form (curve). In my conventions, the
point locus equation of $S=C_2$ reads
  $$
  s_x^2=x^2+y^2+z^2=r_1r^1+r_{-1}r^{-1}+r_0r^0=2r_1r_{-1}-r_0^2=0
  $$
and its envelope equation, $\Si=K_2$,
  $$
  u_\be^2=u^2+v^2+w^2=u_1u^1+u_{-1}u^{-1}+u_0u^0=2u_1u_{-1}-u_0^2=0
  $$
where I have given both the Cartesian and the spherical forms. The latter
is the preferred (canonical) form in analytical geometry\footnote{ \eg\
Salmon, [\pref{Salmon}], Sommerville, [\pref{Sommerville}], Todd,
[\pref{Todd}] \S3.3, Meyer, [\pref{Meyerw}], Grace and Young,
[\pref{GandY}], Semple and Kneebone, [\pref{SandK}].} and what we have
referred to as spherical coordinates are trilinear coordinates with
respect to a specially chosen triangle. For my normalisations, the locus
and envelope equations have exactly the same form.\footnote{ It
corresponds to the choice of $k=2$ in [\pref{SandK}], Chap.V case 4. A
treatment that uses the general form for the base conic is given by
Lindemann, [\pref{Lindemann}]. It is interesting historically to note the
polite way these authors claim priority. This reflects the importance of
invariant theory at that time.} The replacement ${\bf u}\to\nabla$ can be
made in any coordinate system and we have for the polar,
  $$
  h={1\over n(n-1)}\,\De\,a_x^n\equiv\De Y_n({\bf r})=0\,,
  $$
since $Y_n$ is assumed harmonic. This is the announced result whose
significance is that the construction of curves apolar to a conic is a
known topic in invariant theory.

Using suitable parametric coordinates, the analysis of such forms can be
reduced to that of binary forms. The equation of the curve, $C_n$,
factorises into two binary forms of order $n$, Grace and Young,
[\pref{GandY}] \S\S\ 241-244, Schlesinger, [\pref{Schlesinger}]. I give a
discussion since the material is not generally familiar.

The parametric coordinates of a (complex) point, $\ka$, on the
fundamental conic, $S$, are just the Cartan map,
  $$
  r_1=\ka_1^2\,,\quad r_0=\sqrt2\,\ka_1\ka_2,\,\quad r_{-1}=\ka_2^2\,.
  \eql{cparam}
  $$
(I use the notation of [\pref{GandY}] and [\pref{Schlesinger}]. To
compare with (\peq{cm}), $\xi=\ka_1$ and $\eta=\ka_2$.)

If the straight line $u^i\,r_i=0$ cuts the conic in the points $\la$ and
$\mu$ it is easily shown that,
  $$
  u^1=\la_2\mu_2\,,\quad u^0=-{1\over\sqrt2}\,(\la_1\mu_2+\la_2\mu_1),\,
  \quad u^{-1}=\la_1\mu_1\,.
  \eql{param2}
  $$
The tangents at $\la$ and $\mu$ meet at the pole point,
  $$
  r_1=\la_1\mu_1\,,\quad r_0={1\over\sqrt2}\,(\la_1\mu_2+\la_2\mu_1),\,
  \quad r_{-1}=\la_2\mu_2\,.
  \eql{param1}
  $$

Using raising and lowering, one might dispense with the $u$ symbol, since
$ u^i=r^i $, the dual with respect to the conic.\footnote{ This standard
way of representing a point and a line by a pole and its polar with
respect to a fixed conic seems to date back to Hesse in the 1850's
following Poncelet, and others. It was used by Darboux and there exists a
a well known higher dimensional (spin) generalisation due to Clifford,
involving rational norm curves.}

Equation (\peq{param1}) can be written as in (\peq{pauli}) but this time
with two two--spinors, $\la$ and $\mu$,
  $$
   r_i={1\over i\sqrt2}\,\overline\la\,\si_i\,\mu\,.
  $$

If one requires the Cartesian coordinates to be real, then
$\la_1=-\mu_2^*$ and $\la_2=\mu_1^*$ which means that
$\ol\la=\widetilde\mu^*$, \ie\ the complex conjugate transpose, which is
what is normally referred to as the `adjoint', or hermitian conjugate,
$\mu^{{\dag}}$.

Schlesinger gives an instructive way of finding the equation of an
envelope of class $n$ apolar to the base conic, which, as I have said,
corresponds to finding a harmonic homogeneous form. This condition is
equivalent to the envelope being conjugate to all curves of order $n$
that contain $S$ as a factor.

Let $c_x^n$ be such a curve, then the necessary and sufficient condition
that it contain $S$ as a part is that, from (\peq{cparam}),
  $$
  (c^1\ka_1^2+c^0\sqrt2\ka_1\ka_2+c^{-1}\ka_2^2)^n=0
  \eql{factor}
  $$
for each $\ka_1$ and $\ka_2$. The trick now is to think of the symbolic
coefficients, ${\bf c}$, as variables, ${\bf u}$, and $\ka_1$ and $\ka_2$
as fixed {\it coefficient} quantities, in a way to be explained shortly.
The resulting class curve (envelope) will be conjugate to $c_x^n$ by
virtue of (\peq{factor}), with ${\bf c}\to{\bf u}$, and hence will be
apolar to $S$.

Thus we substitute, from (\peq{param2}),
  $$\eqalign{
  c^{1}&=\la_2\mu_2\equiv(\la\mu)_1\cr
  c^0&=-{1\over\sqrt2}\,(\la_1\mu_2+\la_2\mu_1)\equiv(\la\mu)_0\cr
  c^{-1}&=\la_1\mu_1\equiv(\la\mu)_{-1}
  }
  $$
into (\peq{factor}) which factorises into two,
  $$
  (\la_2\ka_1-\la_1\ka_2)^n\,(\mu_2\ka_1-\mu_1\ka_2)^n
  \eql{fact}
  $$
with $\la$ and $\mu$ as variables.

The important point is that, since the $\ka$ are fixed, arbitrary
parameters, one obtains an apolar curve if one replaces the different
products of the $\ka$ by {\it any} quantities. In other words, {\it
$\ka_1$ and $\ka_2$ can be replaced by the symbolic coefficients of a
binary form} so that the above factors become (part) binary forms.
Putting, precisely, $\ka_1=a_2$ and $\ka_2=-a_1$ produces the polarised
binary form,
  $$
  a_\la^n\,a_\mu^n=0\,,
  \eql{acurve}
  $$
as the envelope equation of a curve of the $n$th class apolar to the
fundamental conic, $S$, as soon as one sets $u^i=(\la\mu)_i$, using the
symbolic,
  $$
  a_\la^n\,a_\mu^n=\big[a_1^2(\la\mu)_{-1}-\sqrt2\,a_1a_2(\la\mu)_0
  +a_2^2(\la\mu)_1\big]^n\,.
  \eql{acurve2}
  $$

Precisely the same equation holds for the locus of the apolar curve, but
now the $\la$ and $\mu$ are translated into point coordinates by the dual
(\peq{param1}).

When $\la=\mu=\ka$, the equation, $a_\ka^{2n}=0$, gives the points of
contact of the $2n$ common tangents of the apolar curve and the conic. On
the other hand, the locus equation gives the points where the curve
intersects the conic. The expression in (\peq{acurve}) is the $n$th polar
with respect to $\la$ of  $a_\mu^{2n}$.

What this result says is that a homogeneous harmonic polynomial in
$x,y,z$ can be reconstituted from its evaluation on the absolute conic
$S:{\bf r}^2=0$. I will check this by performing a round trip from the
simple harmonic polynomial, $xy+yz+zx$, through its evaluation on the
absolute to give a binary quartic, thence by polarisation to a product of
two binary quadratics and finally back to $xy+yz+zx$.

For the purposes of solution, it is useful to introduce the
non--homogeneous ratios (the Darboux coordinates),
  $$
  \mu_1:\mu_2=t:1\,,\quad \la_1:\la_2=s:1\,
  $$
so that, \eg,
  $$
  r_1:r_0:r_{-1}=s\,t:{1\over\sqrt2}\,( s+t):1\,.
  \eql{xparams}
  $$

On the absolute, $S$, $s=t$ and (\peq{cparam}) holds. Therefore, on $S$,
  $$
  xy={(t^2+1)(t^2-1)\over2i}\,,\quad yz=it(t^2+1)\,,\quad zx=t(t^2-1)
  $$
and so
  $$
  (xy+yz+zx)\big|_S={1\over2i}\,t^4+(1+i)t^3-(1-i)t-{1\over2i}\,,
  \eql{abin}
  $$
which is a binary quartic. Polarising this according to (\peq{acurve})
one finds, up to a factor,
  $$
  \,t^2\,s^2-(1-i)\,s\,t\,(s+t)-(1+i)\,(t+s)-1\,,
  $$
and substituting (\peq{xparams}) one regains $xy+yz+zx$.

The solutions for the points on $S$ follow by rewriting (\peq{abin})\.,
  $$
  (2\ze^2-2\ze-1)^2=0\,,\quad {\rm with}\quad t=(1-i)\ze\,,
  $$
so that
  $$
  \ze={1\over2}\big(1\pm\sqrt3\big)\,,
  $$
indicating two double roots which are switched under $t\to-1/t^*$. The
corresponding Maxwell poles are proportional to $(1,1,1)$, as we know,
which could have been deduced immediately from the trivial identity,
  $$
  (xy+yz+zx)+{1\over2}\,(x^2+y^2+z^2)={1\over2}\,(x+y+z)^2\,.
  $$

There are three resolutions of
  $$
  (xy+yz+zx)-\la\,(x^2+y^2+z^2)
  \eql{tquad}
  $$
into two linear factors, $\la$ being a solution of the resolving cubic,
  $
  4\la^3-I\la+J=0\,,
  $
of the binary quartic, (\peq{abin}), corresponding (in this
case\footnote{ The general quartic is discussed by Burnside,
[\pref{Burnside}], who gives some geometrical interpretation.}) to
$xy+yz+zx$ via the Cartan map. $I$ and $J$ are the standard invariants of
this quartic, here, $I=3,\,J=-1$. The real resolution occurs for
$\la=-1/2$, (twice since $I^3=27J^2$). The complex one, for $\la=1$, is
exhibited in (\peq{Clebsch2}).

Schlesinger calls the unique apolar curve determined by the binary form,
the curve {\it associated with} the binary form. Thus, given an {\it
arbitrary} curve of the $n$th order, $f_x^n=0$, there is associated with
it, via its $2n$ intersections with the conic, $S$, a unique $n$th order
curve, $a_\la^n\,a_\mu^n=0$, apolar to $S$. This gives a geometrical
interpretation of harmonic projection and of its uniqueness. Furthermore,
there are, in addition to the $2n$ points of intersection, on $S$, of
$a_\la^n\,a_\mu^n=0$ and $f_x^n=0$, another $n(n-2)$ through which it is
known\footnote{ This is Noether's intersection theorem. Roughly speaking,
a curve, $F$, order $n$, that goes through the intersections of curve,
$C$, order $n$, and conic, $S$, order 2, is $F=\rho\,C+MS$, where M is a
curve of order $n-2$ which goes through the remaining intersections of
$F$ and $C$.} that a curve, $m_x^{n-2}=0$, of order $(n-2)$ can be drawn,
that is,
  $$
  f_x^n=\rho\, a_\la^n\,a_\mu^n+\,s^2_x\,m_x^{n-2}\,,
  \eql{fharmp}
  $$
which is the equation expressing the harmonic projection $f\to a\,a$
(with (\peq{param1}) to relate $x$ and $\la,\mu$), equivalent to Gauss'
formula, (\peq{Gauss2}).

A system of $n$ lines, $L_x^n$, through\mgn{GandY.\S247.} the $2n$ points
on $S$ can be considered to be a curve of the $n$th order (an {\it
n--side}) and could be used as a specific $f_x^n$ in (\peq{fharmp}). The
reverse question is whether, given a form apolar to $S$, one can find a
$L_x^n$ such that
  $$
   a_\la^n\,a_\mu^n=\rho\, L_x^n+\,s^2_x\,m_x^{n-2}\,,
  \eql{fharmp2}
  $$
but this would seem to be obvious because the apolar ternary form,
$a_\la^n\,a_\mu^n$, determines the binary form $a_\ka^{2n}$ (and {\it
vice versa}) which fixes the $2n$ points on $S$ from which the $n$--side,
$L_x^n$, can be constructed. Then (\peq{fharmp}) can be employed giving
(\peq{fharmp2}).\footnote{ This way of replacing a binary form by two
ternary ones is outlined in Salmon, [\pref{Salmon}], \S190. See also
Burnside [\pref{Burnside}].} Of course, this is exactly Sylvester's
approach to Maxwell's poles, which is equivalent to Clebsch's earlier
theory.

There are many theorems regarding this set up. For example, it is easy to
show that if two binary forms are conjugate, then the associated curve,
\ie\ harmonic ternary form, of one of them is conjugate to {\it any}
curve, \ie\ ternary form,  passing through the 2$n$ points (on $S$)
corresponding to the other. Conversely, if, through $2n$ points on $S$,
one can draw a curve, $b^n_x=0$, that is conjugate to the apolar curve,
$a_\la^n\,a_\mu^n=0$, then every $n$th order curve that can be drawn
through these points has the same property and the binary form of these
$2n$ points is conjugate to $a_\ka^{2n}$.

For, if $c_x^n=0$ is a second curve which goes through the $2n$ points,
then
  $$
  c_x^n=\rho\,b_x^n+\,s^2_x\,m_x^{n-2}\,.
  $$
However, both curves on the right--hand side are conjugate to
$a_\la^n\,a_\mu^n=0$, the first by assumption and the second because
$a_\la^n\,a_\mu^n=0$ is apolar to $s_x^2=0$ and so, therefore, is
$c_x^n$. Thus a system of $n$ lines containing the $2n$ points represents
a conjugate $n$--side of $a_\la^n\,a_\mu^n=0$ which cuts the conic in
$2n$ points whose binary form is conjugate to $a_\ka^{2n}$.

In terms of $s$ and $t$, the conjugate transformation $\la_1\to-\mu_2^*$,
$\la_2\to \mu_1^*$ is $t\to -1/s^*$. The point coordinates are real (in
Cartesians) if $t\,s^*=-1$ and there are 2 real degrees of freedom, as
required for the real plane.

For applications, one has to impose the condition that the curve be a
real one, \ie\ that the form, (\peq{acurve2}), that gives it is real.
\section{\bf 10. More angular momentum theory.}

After this brief foray into the realms of invariant theory and simple
analytical geometry, I return to angular momentum theory by remarking
that in this, one concentrates on the {\it coefficients}, $\phi^m$, of
the forms, see (\peq{coeffc}), as exemplified in my earlier discussion of
apolarity using $3j$--symbols. Writing the harmonic as (\peq{phi1})
allows one, sometimes, to exploit the considerable amount of existing
angular momentum algebra.

In dealing with higher--spin quantities, which is what $\Phi$ is, there
is always the option of using a (multi) two--spinor description. Since
two--spinors are the defining representation there are few basic
algebraic operations, which is a big advantage, but calculations can
become longwinded. Penrose's, [\pref{Penrose}], treatment of the
algebraic structure of the Riemann curvature is, however, a good example
of the elegance that can be achieved (Pirani, [\pref{Pirani}]). A well
known alternative, for integer spins, is to use `bivector', or
$3$--spinor, space, in which higher spins are represented by symmetric,
traceless tensors, \eg\ $\phi_{m_1\ldots m_j}$, ($m_i=1,0,-1$).\footnote{
This is employed by Debever, Synge and others in the classification of
the Riemann tensor, $j=2$. (See [\pref{RandD}] for comments and a
$(2j+1)$ treatment with the binary quartic concomitant basis written in
3$j$ form.)}

In this connection, I firstly note that the null $2j+1$--spinor,
(\peq{nspin}), which comprises the variables of the binary form, can be
written using $3j$--symbols as, [\pref{DowkandG}],
  $$
  \xi^{(j)}_m={1\over[(2j)!]^{1/2}}\,u^{m_1}(j)\ldots
 u^{m_{2j}}(\textstyle{1\over2})\,\xi^{1\over2}_{m_1}\ldots\xi^{1\over2}_{m_{2j}}\,,
 \eql{fierz1}
  $$
where $u^m$ is the rectangular matrix--spinor exhibited, up to a factor,
as a $3j$--symbol in (\peq{combl}) and
$\xi^{1\over2}=\big({\xi\atop\eta}\big)$. The combination of the $u$s is
the generalised $3j$ symbol that uniquely relates the symmetrised
$\half\otimes\ldots\otimes\half$ representation to the spin-$j$ one (\eg\
Fierz, [\pref{Fierz}], Williams, [\pref{Williams}], Ansari,
[\pref{Ansari}], [\pref{BandL}], pp.421+.]). Assuming $j$ integral, the
two--spinors can be combined in pairs to give the same spin--1 null
vector, $\goa$, ((\peq{cm}), (\peq{pauli})), and, after some $3j$
recombinations, one encounters the familiar polynomial expression for the
spherical harmonics,
  $$\eqalign{
  C^L_M(\goa)=N\Threej L{m_1}mM1{L-1}\Threej {L-1}{m_2}{m'}m1{L-2}\ldots
  \Threej1{m_L}0{m''}10\goa_{m_1}\goa_{m_2}\ldots\goa_{m_L}
  }
  \eql{fierz2}
  $$
where
  $$
  N=\prod_{k=1}^L\bigg[{4k^2-1\over k}\bigg]^{1/2}\,,
  $$
or, in matrix form,
  $$
  C^L(\goa)= N\,A^{m_1}(L)\,A^{m_2}(L-1)\ldots A^{m_L}(1)\,
  \goa_{m_1}\goa_{m_2}\ldots\goa_{m_L}
  \eql{fierz3}
  $$
with the definition,
  $$
  \big[A^{m_1}(L)\big]_{Mm}= \Threej L{m_1}mM1{L-1}\,.
  $$

This expansion is valid for any 3-vector (a 3-spinor) as it follows from
the general recursion,
  $$
  C^L_M({\bf r})=\bigg[{4L^2-1\over L}\bigg]^{1/2}
  \Threej L{m_1}{m_2}M1{L-1}\,
  C^{L-1}_{m_2}({\bf r})\,C^1_{m_1}({\bf r})\,.
  \eql{recurs2}
  $$
which yields,
  $$
  C^L({\bf r})= N\,A^{m_1}(L)\,A^{m_2}(L-1)\ldots A^{m_L}(1)\,
  {\bf r}_{m_1}{\bf r}_{m_2}\ldots{\bf r}_{m_L}\,.
  \eql{fierz5}
  $$

However the `inverse' formula,
  $$
  \goa_{m_1}\goa_{m_2}\ldots\goa_{m_L}
  =C^L(\goa)\,.\,A_{m_1}(L)\,A_{m_2}(L-1)\ldots A_{m_L}(1)\,,
  \eql{fierz4}
  $$
is not valid for any $3$--vector, ${\bf r}$, differing by terms having
${\bf r}^2$ as a factor.

The complete harmonic, (\peq{phi1}), is
  $$\eqalign{
  \Phi_L({\bf r})&=\phi^L\,.\,C^L({\bf r})\cr
  &=N\,\phi^L\,.\,A^{m_1}(L)\,A^{m_2}(L-1)\ldots A^{m_L}(1)\,
  {\bf r}_{m_1}{\bf r}_{m_2}\ldots{\bf r}_{m_L}\cr
  \noalign{\vskip5truept}
  &=\phi^{m_1\ldots m_L}\, {\bf r}_{m_1}{\bf r}_{m_2}\ldots{\bf
  r}_{m_L}\,,
   }
  \eql{comph}
  $$
which defines the symmetric, traceless components, $\phi^{m_1\ldots
m_L}$, of the harmonic in textbook fashion. It is easily checked, using
angular momentum theory, that the product of matrix--spinors,
$A^{m_1}\ldots A^{m_L}$, is traceless on the $m$s.

In this approach, Maxwell's poles are introduced by noting that any
$(2L+1)$--spinor can be expressed in terms of a symmetrised product of,
generally different, $3$--spinors, $\psi^i$,
  $$
   \phi^L=A^{m_1}(L)\,A^{m_2}(L-1)\ldots
  A^{m_L}(1)\,\psi^1_{(m_1}\psi^2_{m_2}\ldots\psi^L_{m_L)}\,,
  \eql{factors}
  $$
corresponding to the irreducible bivector (tensor) equation,
  $$
  \phi_{m_1\ldots m_j}=\psi^1_{(m_1}\psi^2_{m_2}\ldots\psi^L_{m_L)}-{\rm
  trace}\,{\rm terms}\,.
  \eql{factors2}
  $$
(There are no `trace terms' for 2--spinors.)

The reality condition (\peq{cond2}) can be achieved, courtesy of $3j$
properties, by making all the `factors', $\psi^i_m$, obey the same
requirement, \ie\ they correspond to real $3$--vectors.

The factors, $\psi^i$, are the Maxwell poles, ${\bf p}_{(i)}$, as can be
seen by reconstructing the harmonic (\peq{phi1}), using (\peq{comph}) and
(\peq{factors2}),\footnote{ The tensor approach to Maxwell's poles has
been discussed by Zou and Zheng, [\pref{ZandZ}]. See also Applequist,
[\pref{Applequist}].}
  $$
  \Phi_L({\bf r})=\phi^L\,.\,C^L({\bf r})=C\,\prod_{i=1}^L({\bf r}\,.\,\bpsi^i)
  +{\bf r}^2\,G_{L-2}({\bf r})\,,
  \eql{harmm}
  $$
where $G$ arises from the trace terms and can be written out explicitly
in terms of the $\bpsi^i={\bf p}_{(i)}$.

We thus regain the decomposition structure, (\peq{decomp}), in a slightly
heavy--handed, but interesting, way.

A numerical example is always helpful and I return to the harmonic
polynomial used at the end of \S5. The four poles can be found in
MacMillan, [\pref{MacMillan}]. They are, in Cartesians,
  $$
   \eqalign{
  {\bf p}_{(1)}&=(0,0,1)\,,\quad {\bf
  p}_{(2)}={1\over3}\,(0,2\sqrt2,-1)\cr
  {\bf p}_{(3)}&=-{1\over3}(\sqrt6,\sqrt2,1)\,,\quad{\bf p}_{(4)}
  ={1\over3}(\sqrt6,-\sqrt2,-1)\,.
  }
  $$
Then
  $$
  135\prod_{i=1}^4({\bf r}\,.\,{\bf p}_{(i)})=
  -5z\big(12\sqrt2\,x^2y-6x^2z-4\sqrt2\,y^3-6y^2z+z^3\big)\,,
  $$
where the constant, $135$, follows by substituting in values satisfying
$x^2+y^2+z^2=0$. As mentioned before, the last term, $G$, in
(\peq{harmm}) could be found by subtraction but now I wish to obtain it
directly from the poles via the trace terms in (\peq{factors2}) which I
contract with ${\bf r}$s to get
  $$
  r^{m_1}\ldots r^{m_4}\phi_{m_1\ldots m_4}
  =\prod_{i=1}^4({\bf r}\,.\,{\bf p}^{(i)})
-{\rm  trace}\,{\rm terms}\,.
  \eql{factors3}
  $$
Using the standard expression for the traceless part of a fourth rank
tensor\footnote{ Applied either to $\phi$ or to the product of ${\bf
r}$s. A general expression is available, but I will not write it out.}
(\eg\ Jari\'c, [\pref{Jaric}]), or from a short direct calculation, it
follows that,
  $$
  {\rm  trace}\,{\rm terms}={{\bf r}^2\over n+4}\,\Si_6-
  {({\bf r}^2)^2\over (n+4)(n+2)}\,\Si_3\,,\quad n=3\,,
  $$
where $\Si_6$ and $\Si_3$ are the sums of the 6 pair traces and of the 3
double pair traces, respectively,
  $$
  \Si_6={\bf r}^2\,\big({\bf p}_{(3)}\,.\,{\bf p}_{(4)}\big)+\ldots\,,\quad
  \Si_3=\big({\bf p}_{(1)}\,.\,{\bf p}_{(2)}\big)\big({\bf p}_{(3)}\,.\,{\bf
  p}_{(4)}\big)+\ldots\,.
  $$
For my example, arithmetic gives $\Si_6=2{\bf r}^2/9$, $\Si_3=1/3$ and,
simply,
  $$
  G_2={1\over3}\,{\bf r}^2\,,
  $$
agreeing with the result of subtraction.

The fact that the remainder term in (\peq{harmm}) can be expressed in
terms of the poles also follows by noting that there are $L(L-1)/2$
constants in $G_{L-2}$ which could be found by substituting the
$L(L-1)/2$ cross products, ${\bf p}_{(i)}\times{\bf p}_{(j)}$, for ${\bf
r}$ in (\peq{harmm}), when the first term on the right--hand side
vanishes.

The standard spherical harmonic, $C^L_M({\bf r})$, can be expressed in
terms of the null $(2j+1)$--spinors constructed from the $\la$ and $\mu$
introduced earlier, (\peq{param1}),
  $$
  C^{2j}_M({\bf r})=(-1)^{2j}{[(4j+1)!]^{1/2}\over 2^j(2j)!}\,
   \Threej {2j}{m_1}{m_2}M{j}{j}
  \,\la_{m_1}^{(j)}\,\mu_{m_2}^{(j)}\,,\quad 2j=L\,,
  \eql{c3j}
  $$
with $\la^{(\half)}=({\la_1\atop\la_2})$ and
$\mu^{(\half)}=({\mu_1\atop\mu_2})$. Also,
  $$
  (\goa\,.\,{\bf
  r})^{2j}=(\ol\la^{(j)}\,\xi^{(j)})\,(\ol\mu^{(j)}\,\xi^{(j)})=
  \big( (\la_2\,\xi_1-\la_1\,\xi_2)\,\,(\mu_2\,\xi_1-\mu_1\,\xi_2)\big)^{2j}\,,
  \eql{pow}
  $$
(\cf\ (\peq{fact})).

A two--spinor approach to Maxwell's poles and Sylvester's theorem is
given by Dennis, [\pref{Dennis}], who makes some useful comments on the
relation to other techniques.

\section{\bf 11. Discussion.}
Maxwell's motivation for introducing poles was to provide a coordinate
independent geometrical characterisation of spherical harmonics in
3--space; in his own words (in the first edition of his book \footnote{
It is interesting to compare the treatments of this topic in the first
two editions.}) ``emancipating our ideas from the thralldom of systems of
coordinates''. This aspect, no doubt, accounts for the recent rediscovery
of the procedure when analysing astronomical data.

The geometrical viewpoint can also be used advantageously when finding
the harmonics invariant under some finite subgroup of the rotation group.
Maxwell's theory of poles has been employed in this regard by Poole,
[\pref{Poole}], and Laporte, [\pref{Laporte}], although it is not
required of course. Polyhedral harmonics had earlier been introduced by
Pockels in his classic book, [\pref{Pockels}].

It is not clear to me whether the equivalence of apolarity and
harmonicity discussed in \S9 meets Sylvester's requirement of a
geometrical or algebraical definition of a (spherical) harmonic. In fact
it is not obvious that one needs such a definition in view of Maxwell's
rebuttal on pp.201--202 of his second edition (without mentioning names)
of Sylvester's objections.

The limitation to {\it integral} polynomials can be removed. Gauss'
harmonic expansion then becomes an infinite series, as noted by Hobson,
[\pref{Hobson}], p.128. In invariant theory this can be covered by using
Sylvester's {\it perpetuant} construction, which allows the order to
become infinite. \vglue 0.5in

 \noin{\bf References.} \vskip5truept
\begin{putreferences}
  \ref{Heine}{Heine, E. {\it Handbuch der Kugelfunctionen}
  (G.Reimer, Berlin. 1878, 1881).}
  \ref{Pockels}{Pockels, F. {\it \"Uber die Differentialgleichung $\De
  u+k^2u=0$} (Teubner, Leipzig. 1891).}
  \ref{Hamermesh}{Hamermesh, M., {\it Group Theory} (Addison--Wesley,
  Reading. 1962).}
  \ref{Racah}{Racah, G. {\it Group Theory and Spectroscopy}
  (Princeton Lecture Notes, 1951). }
  \ref{Gourdin}{Gourdin, M. {\it Basics of Lie Groups} (Editions
  Fronti\'eres, Gif sur Yvette. 1982.)}
  \ref{Clifford}{Clifford, W.K. \plms{2}{1866}{116}.}
  \ref{Story2}{Story, W.E. \plms{23}{1892}{265}.}
  \ref{Story}{Story, W.E. \ma{41}{1893}{469}.}
  \ref{Poole}{Poole, E.G.C. \plms{33}{1932}{435}.}
  \ref{Dickson}{Dickson, L.E. {\it Algebraic Invariants} (Wiley, N.Y.
  1915).}
  \ref{Dickson2}{Dickson, L.E. {\it Modern Algebraic Theories}
  (Sanborn and Co., Boston. 1926).}
  \ref{Hilbert2}{Hilbert, D. {\it Theory of algebraic invariants} (C.U.P.,
  Cambridge. 1993).}
  \ref{Olver}{Olver, P.J. {\it Classical Invariant Theory} (C.U.P., Cambridge.
  1999.)}
  \ref{AST}{A\v{s}erova, R.M., Smirnov, J.F. and Tolsto\v{i}, V.N. {\it
  Teoret. Mat. Fyz.} {\bf 8} (1971) 255.}
  \ref{AandS}{A\v{s}erova, R.M., Smirnov, J.F. \np{4}{1968}{399}.}
  \ref{Shapiro}{Shapiro, J. \jmp{6}{1965}{1680}.}
  \ref{Shapiro2}{Shapiro, J.Y. \jmp{14}{1973}{1262}.}
  \ref{NandS}{Noz, M.E. and Shapiro, J.Y. \np{51}{1973}{309}.}
  \ref{Cayley2}{Cayley, A. {\it Phil. Trans. Roy. Soc. Lond.}
  {\bf 144} (1854) 244.}
  \ref{Cayley3}{Cayley, A. {\it Phil. Trans. Roy. Soc. Lond.}
  {\bf 146} (1856) 101.}
  \ref{Wigner}{Wigner, E.P. {\it Gruppentheorie} (Vieweg, Braunschweig. 1931).}
  \ref{Sharp}{Sharp, R.T. \ajop{28}{1960}{116}.}
  \ref{Laporte}{Laporte, O. {\it Z. f. Naturf.} {\bf 3a} (1948) 447.}
  \ref{Lowdin}{L\"owdin, P-O. \rmp{36}{1964}{966}.}
  \ref{Ansari}{Ansari, S.M.R. {\it Fort. d. Phys.} {\bf 15} (1967) 707.}
  \ref{SSJR}{Samal, P.K., Saha, R., Jain, P. and Ralston, J.P. {\it
  Testing Isotropy of Cosmic Microwave Background Radiation},
  astro-ph/0708.2816.}
  \ref{Lachieze}{Lachi\'eze-Rey, M. {\it Harmonic projection and
  multipole Vectors}. astro- \break ph/0409081.}
  \ref{CHS}{Copi, C.J., Huterer, D. and Starkman, G.D.
  \prD{70}{2003}{043515}.}
  \ref{Jaric}{Jari\'c, J.P. {\it Int. J. Eng. Sci.} {\bf 41} (2003) 2123.}
  \ref{RandD}{Roche, J.A. and Dowker, J.S. \jpa{1}{1968}{527}.}
  \ref{KandW}{Katz, G. and Weeks, J.R. \prD{70}{2004}{063527}.}
  \ref{Waerden}{van der Waerden, B.L. {\it Die Gruppen-theoretische
  Methode in der Quantenmechanik} (Springer, Berlin. 1932).}
  \ref{EMOT}{Erdelyi, A., Magnus, W., Oberhettinger, F. and Tricomi, F.G. {
  \it Higher Transcendental Functions} Vol.1 (McGraw-Hill, N.Y. 1953).}
  \ref{Dowzilch}{Dowker, J.S. {\it Proc. Phys. Soc.} {\bf 91} (1967) 28.}
  \ref{DandD}{Dowker, J.S. and Dowker, Y.P. {\it Proc. Phys. Soc.}
  {\bf 87} (1966) 65.}
  \ref{CoandH}{Courant, R. and Hilbert, D. {\it Methoden der
  Mathematischen Physik} vol.1 \break (Springer, Berlin. 1931).}
  \ref{Applequist}{Applequist, J. \jpa{22}{1989}{4303}.}
  \ref{Torruella}{Torruella, \jmp{16}{1975}{1637}.}
  \ref{Weinberg}{Weinberg, S.W. \pr{133}{1964}{B1318}.}
  \ref{Meyerw}{Meyer, W.F. {\it Apolarit\"at und rationale Curven}
  (Fues, T\"ubingen. 1883.) }
  \ref{Ostrowski}{Ostrowski, A. {\it Jahrsb. Deutsch. Math. Verein.} {\bf
  33} (1923) 245.}
  \ref{Kramers}{Kramers, H.A. {\it Grundlagen der Quantenmechanik}, (Akad.
  Verlag., Leipzig, 1938).}
  \ref{ZandZ}{Zou, W.-N. and Zheng, Q.-S. \prs{459}{2003}{527}.}
  \ref{Weeks1}{Weeks, J.R. {\it Maxwell's multipole vectors
  and the CMB}.  astro-ph/0412231.}
  \ref{Corson}{Corson, E.M. {\it Tensors, Spinors and Relativistic Wave
  Equations} (Blackie, London. 1950).}
  \ref{Rosanes}{Rosanes, J. \jram{76}{1873}{312}.}
  \ref{Salmon}{Salmon, G. {\it Lessons Introductory to the Modern Higher
  Algebra} 3rd. edn. \break (Hodges,  Dublin. 1876.)}
  \ref{Milnew}{Milne, W.P. {\it Homogeneous Coordinates} (Arnold. London. 1910).}
  \ref{Niven}{Niven, W.D. {\it Phil. Trans. Roy. Soc.} {\bf 170} (1879) 393.}
  \ref{Scott}{Scott, C.A. {\it An Introductory Account of
  Certain Modern Ideas and Methods in Plane Analytical Geometry,}
  (MacMillan, N.Y. 1896).}
  \ref{Bargmann}{Bargmann, V. \rmp{34}{1962}{300}.}
  \ref{Maxwell}{Maxwell, J.C. {\it A Treatise on Electricity and
  Magnetism} 2nd. edn. (Clarendon Press, Oxford. 1882).}
  \ref{BandL}{Biedenharn, L.C. and Louck, J.D. {\it Angular Momentum in Quantum Physics}
  (Addison-Wesley, Reading. 1981).}
  \ref{Weylqm}{Weyl, H. {\it The Theory of Groups and Quantum Mechanics}
  (Methuen, London. 1931).}
  \ref{Robson}{Robson, A. {\it An Introduction to Analytical Geometry} Vol I
  (C.U.P., Cambridge. 1940.)}
  \ref{Sommerville}{Sommerville, D.M.Y. {\it Analytical Conics} 3rd. edn.
   (Bell. London. 1933).}
  \ref{Coolidge}{Coolidge, J.L. {\it A Treatise on Algebraic Plane Curves}
  (Clarendon Press, Oxford. 1931).}
  \ref{SandK}{Semple, G. and Kneebone. G.T. {\it Algebraic Projective
  Geometry} (Clarendon Press, Oxford. 1952).}
  \ref{AandC}{Abdesselam A., and Chipalkatti, J. {\it The Higher
  Transvectants are redundant}, arXiv:0801.1533 [math.AG] 2008.}
  \ref{Elliott}{Elliott, E.B. {\it The Algebra of Quantics} 2nd edn.
  (Clarendon Press, Oxford. 1913).}
  \ref{Elliott2}{Elliott, E.B. \qjm{48}{1917}{372}.}
  \ref{Howe}{Howe, R. \tams{313}{1989}{539}.}
  \ref{Clebsch}{Clebsch, A. \jram{60}{1862}{343}.}
  \ref{Prasad}{Prasad, G. \ma{72}{1912}{136}.}
  \ref{Dougall}{Dougall, J. \pems{32}{1913}{30}.}
  \ref{Penrose}{Penrose, R. \aop{10}{1960}{171}.}
  \ref{Penrose2}{Penrose, R. \prs{273}{1965}{171}.}
  \ref{Burnside}{Burnside, W.S. \qjm{10}{1870}{211}. }
  \ref{Lindemann}{Lindemann, F. \ma{23} {1884}{111}.}
  \ref{Backus}{Backus, G. {\it Rev. Geophys. Space Phys.} {\bf 8} (1970) 633.}
  \ref{Baerheim}{Baerheim, R. {\it Q.J. Mech. appl. Math.} {\bf 51} (1998) 73.}
  \ref{Lense}{Lense, J. {\it Kugelfunktionen} (Akad.Verlag, Leipzig. 1950).}
  \ref{Littlewood}{Littlewood, D.E. \plms{50}{1948}{349}.}
  \ref{Fierz}{Fierz, M. {\it Helv. Phys. Acta} {\bf 12} (1938) 3.}
  \ref{Williams}{Williams, D.N. {\it Lectures in Theoretical Physics} Vol. VII,
  (Univ.Colorado Press, Boulder. 1965).}
  \ref{Dennis}{Dennis, M. \jpa{37}{2004}{9487}.}
  \ref{Pirani}{Pirani, F. {\it Brandeis Lecture Notes on
  General Relativity,} edited by S. Deser and K. Ford. (Brandeis, Mass. 1964).}
  \ref{Sturm}{Sturm, R. \jram{86}{1878}{116}.}
  \ref{Schlesinger}{Schlesinger, O. \ma{22}{1883}{521}.}
  \ref{Askwith}{Askwith, E.H. {\it Analytical Geometry of the Conic
  Sections} (A.\&C. Black, London. 1908).}
  \ref{Todd}{Todd, J.A. {\it Projective and Analytical Geometry}.
  (Pitman, London. 1946).}
  \ref{Glenn}{Glenn. O.E. {\it Theory of Invariants} (Ginn \& Co, N.Y. 1915).}
  \ref{DowkandG}{Dowker, J.S. and Goldstone, M. \prs{303}{1968}{381}.}
  \ref{Turnbull}{Turnbull, H.A. {\it The Theory of Determinants,
  Matrices and Invariants} 3rd. edn. (Dover, N.Y. 1960).}
  \ref{MacMillan}{MacMillan, W.D. {\it The Theory of the Potential}
  (McGraw-Hill, N.Y. 1930).}
   \ref{Hobson}{Hobson, E.W. {\it The Theory of Spherical and Ellipsoidal Harmonics}
   C.U.P., Cambridge. 1931).}
  \ref{Hobson1}{Hobson, E.W. \plms {24}{1892}{55}.}
  \ref{GandY}{Grace, J.H. and Young, A. {\it The Algebra of Invariants}
  (C.U.P., Cambridge, 1903).}
  \ref{FandR}{Fano, U. and Racah, G. {\it Irreducible Tensorial Sets}
  (Academic Press, N.Y. 1959).}
  \ref{TandT}{Thomson, W. and Tait, P.G. {\it Treatise on Natural Philosophy}
  (Clarendon Press, Oxford. 1867).}
  \ref{Brinkman}{Brinkman, H.C. {\it Applications of spinor invariants in
atomic physics}, North Holland, Amsterdam 1956.}
  \ref{Kramers1}{Kramers, H.A. {\it Proc. Roy. Soc. Amst.} {\bf 33} (1930) 953.}
  \ref{DandP2}{Dowker,J.S. and Pettengill,D.F. \jpa{7}{1974}{1527}}
  \ref{Dowk1}{Dowker,J.S. \jpa{}{}{45}.}
  \ref{Dowk2}{Dowker,J.S. \aop{71}{1972}{577}}
  \ref{DandA}{Dowker,J.S. and Apps, J.S. \cqg{15}{1998}{1121}.}
  \ref{Weil}{Weil,A., {\it Elliptic functions according to Eisenstein
  and Kronecker}, Springer, Berlin, 1976.}
  \ref{Ling}{Ling,C-H. {\it SIAM J.Math.Anal.} {\bf5} (1974) 551.}
  \ref{Ling2}{Ling,C-H. {\it J.Math.Anal.Appl.}(1988).}
 \ref{BMO}{Brevik,I., Milton,K.A. and Odintsov, S.D. \aop{302}{2002}{120}.}
 \ref{KandL}{Kutasov,D. and Larsen,F. {\it JHEP} 0101 (2001) 1.}
 \ref{KPS}{Klemm,D., Petkou,A.C. and Siopsis {\it Entropy
 bounds, monoticity properties and scaling in CFT's}. hep-th/0101076.}
 \ref{DandC}{Dowker,J.S. and Critchley,R. \prD{15}{1976}{1484}.}
 \ref{AandD}{Al'taie, M.B. and Dowker, J.S. \prD{18}{1978}{3557}.}
 \ref{Dow1}{Dowker,J.S. \prD{37}{1988}{558}.}
 \ref{Dow30}{Dowker,J.S. \prD{28}{1983}{3013}.}
 \ref{DandK}{Dowker,J.S. and Kennedy,G. \jpa{}{1978}{}.}
 \ref{Dow2}{Dowker,J.S. \cqg{1}{1984}{359}.}
 \ref{DandKi}{Dowker,J.S. and Kirsten, K. {\it Comm. in Anal. and Geom.
 }{\bf7} (1999) 641.}
 \ref{DandKe}{Dowker,J.S. and Kennedy,G.\jpa{11}{1978}{895}.}
 \ref{Gibbons}{Gibbons,G.W. \pl{60A}{1977}{385}.}
 \ref{Cardy}{Cardy,J.L. \np{366}{1991}{403}.}
 \ref{ChandD}{Chang,P. and Dowker,J.S. \np{395}{1993}{407}.}
 \ref{DandC2}{Dowker,J.S. and Critchley,R. \prD{13}{1976}{224}.}
 \ref{Camporesi}{Camporesi,R. \prp{196}{1990}{1}.}
 \ref{BandM}{Brown,L.S. and Maclay,G.J. \pr{184}{1969}{1272}.}
 \ref{CandD}{Candelas,P. and Dowker,J.S. \prD{19}{1979}{2902}.}
 \ref{Unwin1}{Unwin,S.D. Thesis. University of Manchester. 1979.}
 \ref{Unwin2}{Unwin,S.D. \jpa{13}{1980}{313}.}
 \ref{DandB}{Dowker,J.S.and Banach,R. \jpa{11}{1978}{2255}.}
 \ref{Obhukov}{Obhukov,Yu.N. \pl{109B}{1982}{195}.}
 \ref{Kennedy}{Kennedy,G. \prD{23}{1981}{2884}.}
 \ref{CandT}{Copeland,E. and Toms,D.J. \np {255}{1985}{201}.}
 \ref{ELV}{Elizalde,E., Lygren, M. and Vassilevich,
 D.V. \jmp {37}{1996}{3105}.}
 \ref{Malurkar}{Malurkar,S.L. {\it J.Ind.Math.Soc} {\bf16} (1925/26) 130.}
 \ref{Glaisher}{Glaisher,J.W.L. {\it Messenger of Math.} {\bf18}
(1889) 1.} \ref{Anderson}{Anderson,A. \prD{37}{1988}{536}.}
 \ref{CandA}{Cappelli,A. and D'Appollonio,\pl{487B}{2000}{87}.}
 \ref{Wot}{Wotzasek,C. \jpa{23}{1990}{1627}.}
 \ref{RandT}{Ravndal,F. and Tollesen,D. \prD{40}{1989}{4191}.}
 \ref{SandT}{Santos,F.C. and Tort,A.C. \pl{482B}{2000}{323}.}
 \ref{FandO}{Fukushima,K. and Ohta,K. {\it Physica} {\bf A299} (2001) 455.}
 \ref{GandP}{Gibbons,G.W. and Perry,M. \prs{358}{1978}{467}.}
 \ref{Dow4}{Dowker,J.S..}
  \ref{Rad}{Rademacher,H. {\it Topics in analytic number theory,}
Springer-Verlag,  Berlin,1973.}
  \ref{Halphen}{Halphen,G.-H. {\it Trait\'e des Fonctions Elliptiques},
  Vol 1, Gauthier-Villars, Paris, 1886.}
  \ref{CandW}{Cahn,R.S. and Wolf,J.A. {\it Comm.Mat.Helv.} {\bf 51}
  (1976) 1.}
  \ref{Berndt}{Berndt,B.C. \rmjm{7}{1977}{147}.}
  \ref{Hurwitz}{Hurwitz,A. \ma{18}{1881}{528}.}
  \ref{Hurwitz2}{Hurwitz,A. {\it Mathematische Werke} Vol.I. Basel,
  Birkhauser, 1932.}
  \ref{Berndt2}{Berndt,B.C. \jram{303/304}{1978}{332}.}
  \ref{RandA}{Rao,M.B. and Ayyar,M.V. \jims{15}{1923/24}{150}.}
  \ref{Hardy}{Hardy,G.H. \jlms{3}{1928}{238}.}
  \ref{TandM}{Tannery,J. and Molk,J. {\it Fonctions Elliptiques},
   Gauthier-Villars, Paris, 1893--1902.}
  \ref{schwarz}{Schwarz,H.-A. {\it Formeln und
  Lehrs\"atzen zum Gebrauche..},Springer 1893.(The first edition was 1885.)
  The French translation by Henri Pad\'e is {\it Formules et Propositions
  pour L'Emploi...},Gauthier-Villars, Paris, 1894}
  \ref{Hancock}{Hancock,H. {\it Theory of elliptic functions}, Vol I.
   Wiley, New York 1910.}
  \ref{watson}{Watson,G.N. \jlms{3}{1928}{216}.}
  \ref{MandO}{Magnus,W. and Oberhettinger,F. {\it Formeln und S\"atze},
  Springer-Verlag, Berlin 1948.}
  \ref{Klein}{Klein,F. {\it Lectures on the Icosohedron}
  (Methuen, London, 1913).}
  \ref{AandL}{Appell,P. and Lacour,E. {\it Fonctions Elliptiques},
  Gauthier-Villars,
  Paris, 1897.}
  \ref{HandC}{Hurwitz,A. and Courant,C. {\it Allgemeine Funktionentheorie},
  Springer,
  Berlin, 1922.}
  \ref{WandW}{Whittaker,E.T. and Watson,G.N. {\it Modern analysis},
  Cambridge 1927.}
  \ref{SandC}{Selberg,A. and Chowla,S. \jram{227}{1967}{86}. }
  \ref{zucker}{Zucker,I.J. {\it Math.Proc.Camb.Phil.Soc} {\bf 82 }(1977)
  111.}
  \ref{glasser}{Glasser,M.L. {\it Maths.of Comp.} {\bf 25} (1971) 533.}
  \ref{GandW}{Glasser, M.L. and Wood,V.E. {\it Maths of Comp.} {\bf 25}
  (1971)
  535.}
  \ref{greenhill}{Greenhill,A,G. {\it The Applications of Elliptic
  Functions}, MacMillan, London, 1892.}
  \ref{Weierstrass}{Weierstrass,K. {\it J.f.Mathematik (Crelle)}
{\bf 52} (1856) 346.}
  \ref{Weierstrass2}{Weierstrass,K. {\it Mathematische Werke} Vol.I,p.1,
  Mayer u. M\"uller, Berlin, 1894.}
  \ref{Fricke}{Fricke,R. {\it Die Elliptische Funktionen und Ihre Anwendungen},
    Teubner, Leipzig. 1915, 1922.}
  \ref{Konig}{K\"onigsberger,L. {\it Vorlesungen \"uber die Theorie der
 Elliptischen Funktionen},  \break Teubner, Leipzig, 1874.}
  \ref{Milne}{Milne,S.C. {\it The Ramanujan Journal} {\bf 6} (2002) 7-149.}
  \ref{Schlomilch}{Schl\"omilch,O. {\it Ber. Verh. K. Sachs. Gesell. Wiss.
  Leipzig}  {\bf 29} (1877) 101-105; {\it Compendium der h\"oheren
  Analysis}, Bd.II, 3rd Edn, Vieweg, Brunswick, 1878.}
  \ref{BandB}{Briot,C. and Bouquet,C. {\it Th\`eorie des Fonctions
  Elliptiques}, Gauthier-Villars, Paris, 1875.}
  \ref{Dumont}{Dumont,D. \aim {41}{1981}{1}.}
  \ref{Andre}{Andr\'e,D. {\it Ann.\'Ecole Normale Superior} {\bf 6} (1877)
  265;
  {\it J.Math.Pures et Appl.} {\bf 5} (1878) 31.}
  \ref{Raman}{Ramanujan,S. {\it Trans.Camb.Phil.Soc.} {\bf 22} (1916) 159;
 {\it Collected Papers}, Cambridge, 1927}
  \ref{Weber}{Weber,H.M. {\it Lehrbuch der Algebra} Bd.III, Vieweg,
  Brunswick 190  3.}
  \ref{Weber2}{Weber,H.M. {\it Elliptische Funktionen und algebraische
  Zahlen},
  Vieweg, Brunswick 1891.}
  \ref{ZandR}{Zucker,I.J. and Robertson,M.M.
  {\it Math.Proc.Camb.Phil.Soc} {\bf 95 }(1984) 5.}
  \ref{JandZ1}{Joyce,G.S. and Zucker,I.J.
  {\it Math.Proc.Camb.Phil.Soc} {\bf 109 }(1991) 257.}
  \ref{JandZ2}{Zucker,I.J. and Joyce.G.S.
  {\it Math.Proc.Camb.Phil.Soc} {\bf 131 }(2001) 309.}
  \ref{zucker2}{Zucker,I.J. {\it SIAM J.Math.Anal.} {\bf 10} (1979) 192,}
  \ref{BandZ}{Borwein,J.M. and Zucker,I.J. {\it IMA J.Math.Anal.} {\bf 12}
  (1992) 519.}
  \ref{Cox}{Cox,D.A. {\it Primes of the form $x^2+n\,y^2$}, Wiley,
  New York, 1989.}
  \ref{BandCh}{Berndt,B.C. and Chan,H.H. {\it Mathematika} {\bf42} (1995)
  278.}
  \ref{EandT}{Elizalde,R. and Tort.hep-th/}
  \ref{KandS}{Kiyek,K. and Schmidt,H. {\it Arch.Math.} {\bf 18} (1967) 438.}
  \ref{Oshima}{Oshima,K. \prD{46}{1992}{4765}.}
  \ref{greenhill2}{Greenhill,A.G. \plms{19} {1888} {301}.}
  \ref{Russell}{Russell,R. \plms{19} {1888} {91}.}
  \ref{BandB}{Borwein,J.M. and Borwein,P.B. {\it Pi and the AGM}, Wiley,
  New York, 1998.}
  \ref{Resnikoff}{Resnikoff,H.L. \tams{124}{1966}{334}.}
  \ref{vandp}{Van der Pol, B. {\it Indag.Math.} {\bf18} (1951) 261,272.}
  \ref{Rankin}{Rankin,R.A. {\it Modular forms} C.U.P. Cambridge}
  \ref{Rankin2}{Rankin,R.A. {\it Proc. Roy.Soc. Edin.} {\bf76 A} (1976) 107.}
  \ref{Skoruppa}{Skoruppa,N-P. {\it J.of Number Th.} {\bf43} (1993) 68 .}
  \ref{Down}{Dowker.J.S. \np {104}{2002}{153}.}
  \ref{Eichler}{Eichler,M. \mz {67}{1957}{267}.}
  \ref{Zagier}{Zagier,D. \invm{104}{1991}{449}.}
  \ref{Lang}{Lang,S. {\it Modular Forms}, Springer, Berlin, 1976.}
  \ref{Kosh}{Koshliakov,N.S. {\it Mess.of Math.} {\bf 58} (1928) 1.}
  \ref{BandH}{Bodendiek, R. and Halbritter,U. \amsh{38}{1972}{147}.}
  \ref{Smart}{Smart,L.R., \pgma{14}{1973}{1}.}
  \ref{Grosswald}{Grosswald,E. {\it Acta. Arith.} {\bf 21} (1972) 25.}
  \ref{Kata}{Katayama,K. {\it Acta Arith.} {\bf 22} (1973) 149.}
  \ref{Ogg}{Ogg,A. {\it Modular forms and Dirichlet series} (Benjamin,
  New York,
   1969).}
  \ref{Bol}{Bol,G. \amsh{16}{1949}{1}.}
  \ref{Epstein}{Epstein,P. \ma{56}{1903}{615}.}
  \ref{Petersson}{Petersson.}
  \ref{Serre}{Serre,J-P. {\it A Course in Arithmetic}, Springer,
  New York, 1973.}
  \ref{Schoenberg}{Schoenberg,B., {\it Elliptic Modular Functions},
  Springer, Berlin, 1974.}
  \ref{Apostol}{Apostol,T.M. \dmj {17}{1950}{147}.}
  \ref{Ogg2}{Ogg,A. {\it Lecture Notes in Math.} {\bf 320} (1973) 1.}
  \ref{Knopp}{Knopp,M.I. \dmj {45}{1978}{47}.}
  \ref{Knopp2}{Knopp,M.I. \invm {}{1994}{361}.}
  \ref{LandZ}{Lewis,J. and Zagier,D. \aom{153}{2001}{191}.}
  \ref{DandK1}{Dowker,J.S. and Kirsten,K. {\it Elliptic functions and
  temperature inversion symmetry on spheres} hep-th/.}
  \ref{HandK}{Husseini and Knopp.}
  \ref{Kober}{Kober,H. \mz{39}{1934-5}{609}.}
  \ref{HandL}{Hardy,G.H. and Littlewood, \am{41}{1917}{119}.}
  \ref{Watson}{Watson,G.N. \qjm{2}{1931}{300}.}
  \ref{SandC2}{Chowla,S. and Selberg,A. {\it Proc.Nat.Acad.} {\bf 35}
  (1949) 371.}
  \ref{Landau}{Landau, E. {\it Lehre von der Verteilung der Primzahlen},
  (Teubner, Leipzig, 1909).}
  \ref{Berndt4}{Berndt,B.C. \tams {146}{1969}{323}.}
  \ref{Berndt3}{Berndt,B.C. \tams {}{}{}.}
  \ref{Bochner}{Bochner,S. \aom{53}{1951}{332}.}
  \ref{Weil2}{Weil,A.\ma{168}{1967}{}.}
  \ref{CandN}{Chandrasekharan,K. and Narasimhan,R. \aom{74}{1961}{1}.}
  \ref{Rankin3}{Rankin,R.A. {} {} ().}
  \ref{Berndt6}{Berndt,B.C. {\it Trans.Edin.Math.Soc}.}
  \ref{Elizalde}{Elizalde,E. {\it Ten Physical Applications of Spectral
  Zeta Function Theory}, \break (Springer, Berlin, 1995).}
  \ref{Allen}{Allen,B., Folacci,A. and Gibbons,G.W. \pl{189}{1987}{304}.}
  \ref{Krazer}{Krazer}
  \ref{Elizalde3}{Elizalde,E. {\it J.Comp.and Appl. Math.} {\bf 118}
  (2000) 125.}
  \ref{Elizalde2}{Elizalde,E., Odintsov.S.D, Romeo, A. and Bytsenko,
  A.A and
  Zerbini,S.
  {\it Zeta function regularisation}, (World Scientific, Singapore,
  1994).}
  \ref{Eisenstein}{Eisenstein}
  \ref{Hecke}{Hecke,E. \ma{112}{1936}{664}.}
  \ref{Terras}{Terras,A. {\it Harmonic analysis on Symmetric Spaces} (Springer,
  New York, 1985).}
  \ref{BandG}{Bateman,P.T. and Grosswald,E. {\it Acta Arith.} {\bf 9}
  (1964) 365.}
  \ref{Deuring}{Deuring,M. \aom{38}{1937}{585}.}
  \ref{Guinand}{Guinand.}
  \ref{Guinand2}{Guinand.}
  \ref{Minak}{Minakshisundaram.}
  \ref{Mordell}{Mordell,J. \prs{}{}{}.}
  \ref{GandZ}{Glasser,M.L. and Zucker, {}.}
  \ref{Landau2}{Landau,E. \jram{}{1903}{64}.}
  \ref{Kirsten1}{Kirsten,K. \jmp{35}{1994}{459}.}
  \ref{Sommer}{Sommer,J. {\it Vorlesungen \"uber Zahlentheorie}
  (1907,Teubner,Leipzig).
  French edition 1913 .}
  \ref{Reid}{Reid,L.W. {\it Theory of Algebraic Numbers},
  (1910,MacMillan,New York).}
  \ref{Milnor}{Milnor, J. {\it Is the Universe simply--connected?},
  IAS, Princeton, 1978.}
  \ref{Milnor2}{Milnor, J. \ajm{79}{1957}{623}.}
  \ref{Opechowski}{Opechowski,W. {\it Physica} {\bf 7} (1940) 552.}
  \ref{Bethe}{Bethe, H.A. \zfp{3}{1929}{133}.}
  \ref{LandL}{Landau, L.D. and Lishitz, E.M. {\it Quantum
  Mechanics} (Pergamon Press, London, 1958).}
  \ref{GPR}{Gibbons, G.W., Pope, C. and R\"omer, H., \np{157}{1979}{377}.}
  \ref{Jadhav}{Jadhav,S.P. PhD Thesis, University of Manchester 1990.}
  \ref{DandJ}{Dowker,J.S. and Jadhav, S. \prD{39}{1989}{1196}.}
  \ref{CandM}{Coxeter, H.S.M. and Moser, W.O.J. {\it Generators and
  relations of finite groups} Springer. Berlin. 1957.}
  \ref{Coxeter2}{Coxeter, H.S.M. {\it Regular Complex Polytopes},
   (Cambridge University Press,
  Cambridge, 1975).}
  \ref{Coxeter}{Coxeter, H.S.M. {\it Regular Polytopes}.}
  \ref{Stiefel}{Stiefel, E., J.Research NBS {\bf 48} (1952) 424.}
  \ref{BandS}{Brink, D.M. and Satchler, G.R. {\it Angular momentum theory}.
  (Clarendon Press, Oxford. 1962.).}
  \ref{Rose}{Rose}
  \ref{Schwinger}{Schwinger, J. {\it On Angular Momentum} in {\it Quantum Theory of
  Angular Momentum} edited by Biedenharn,L.C. and van Dam, H.
  (Academic Press, N.Y. 1965).}
  \ref{Bromwich}{Bromwich, T.J.I'A. {\it Infinite Series},
  (Macmillan, 1947).}
  \ref{Ray}{Ray,D.B. \aim{4}{1970}{109}.}
  \ref{Ikeda}{Ikeda,A. {\it Kodai Math.J.} {\bf 18} (1995) 57.}
  \ref{Kennedy}{Kennedy,G. \prD{23}{1981}{2884}.}
  \ref{Ellis}{Ellis,G.F.R. {\it General Relativity} {\bf2} (1971) 7.}
  \ref{Dow8}{Dowker,J.S. \cqg{20}{2003}{L105}.}
  \ref{IandY}{Ikeda, A and Yamamoto, Y. \ojm {16}{1979}{447}.}
  \ref{BandI}{Bander,M. and Itzykson,C. \rmp{18}{1966}{2}.}
  \ref{Schulman}{Schulman, L.S. \pr{176}{1968}{1558}.}
  \ref{Bar1}{B\"ar,C. {\it Arch.d.Math.}{\bf 59} (1992) 65.}
  \ref{Bar2}{B\"ar,C. {\it Geom. and Func. Anal.} {\bf 6} (1996) 899.}
  \ref{Vilenkin}{Vilenkin, N.J. {\it Special functions},
  (Am.Math.Soc., Providence, 1968).}
  \ref{Talman}{Talman, J.D. {\it Special functions} (Benjamin,N.Y.,1968).}
  \ref{Miller}{Miller, W. {\it Symmetry groups and their applications}
  (Wiley, N.Y., 1972).}
  \ref{Dow3}{Dowker,J.S. \cmp{162}{1994}{633}.}
  \ref{Cheeger}{Cheeger, J. \jdg {18}{1983}{575}.}
  \ref{Dow6}{Dowker,J.S. \jmp{30}{1989}{770}.}
  \ref{Dow20}{Dowker,J.S. \jmp{35}{1994}{6076}.}
  \ref{Dow21}{Dowker,J.S. {\it Heat kernels and polytopes} in {\it
   Heat Kernel Techniques and Quantum Gravity}, ed. by S.A.Fulling,
   Discourses in Mathematics and its Applications, No.4, Dept.
   Maths., Texas A\&M University, College Station, Texas, 1995.}
  \ref{Dow9}{Dowker,J.S. \jmp{42}{2001}{1501}.}
  \ref{Dow7}{Dowker,J.S. \jpa{25}{1992}{2641}.}
  \ref{Warner}{Warner.N.P. \prs{383}{1982}{379}.}
  \ref{Wolf}{Wolf, J.A. {\it Spaces of constant curvature},
  (McGraw--Hill,N.Y., 1967).}
  \ref{Meyer}{Meyer,B. \cjm{6}{1954}{135}.}
  \ref{BandB}{B\'erard,P. and Besson,G. {\it Ann. Inst. Four.} {\bf 30}
  (1980) 237.}
  \ref{PandM}{Polya,G. and Meyer,B. \cras{228}{1948}{28}.}
  \ref{Springer}{Springer, T.A. Lecture Notes in Math. vol 585 (Springer,
  Berlin,1977).}
  \ref{SeandT}{Threlfall, H. and Seifert, W. \ma{104}{1930}{1}.}
  \ref{Hopf}{Hopf,H. \ma{95}{1925}{313}. }
  \ref{Dow}{Dowker,J.S. \jpa{5}{1972}{936}.}
  \ref{LLL}{Lehoucq,R., Lachi\'eze-Rey,M. and Luminet, J.--P. {\it
  Astron.Astrophys.} {\bf 313} (1996) 339.}
  \ref{LaandL}{Lachi\'eze-Rey,M. and Luminet, J.--P.
  \prp{254}{1995}{135}.}
  \ref{Schwarzschild}{Schwarzschild, K., {\it Vierteljahrschrift der
  Ast.Ges.} {\bf 35} (1900) 337.}
  \ref{Starkman}{Starkman,G.D. \cqg{15}{1998}{2529}.}
  \ref{LWUGL}{Lehoucq,R., Weeks,J.R., Uzan,J.P., Gausman, E. and
  Luminet, J.--P. \cqg{19}{2002}{4683}.}
  \ref{Dow10}{Dowker,J.S. \prD{28}{1983}{3013}.}
  \ref{BandD}{Banach, R. and Dowker, J.S. \jpa{12}{1979}{2527}.}
  \ref{Jadhav2}{Jadhav,S. \prD{43}{1991}{2656}.}
  \ref{Gilkey}{Gilkey,P.B. {\it Invariance theory,the heat equation and
  the Atiyah--Singer Index theorem} (CRC Press, Boca Raton, 1994).}
  \ref{BandY}{Berndt,B.C. and Yeap,B.P. {\it Adv. Appl. Math.}
  {\bf29} (2002) 358.}
  \ref{HandR}{Hanson,A.J. and R\"omer,H. \pl{80B}{1978}{58}.}
  \ref{Hill}{Hill,M.J.M. {\it Trans.Camb.Phil.Soc.} {\bf 13} (1883) 36.}
  \ref{Cayley}{Cayley,A. {\it Quart.Math.J.} {\bf 7} (1866) 304.}
  \ref{Seade}{Seade,J.A. {\it Anal.Inst.Mat.Univ.Nac.Aut\'on
  M\'exico} {\bf 21} (1981) 129.}
  \ref{CM}{Cisneros--Molina,J.L. {\it Geom.Dedicata} {\bf84} (2001)
  \ref{Goette1}{Goette,S. \jram {526} {2000} 181.}
  207.}
  \ref{NandO}{Nash,C. and O'Connor,D--J, \jmp {36}{1995}{1462}.}
  \ref{Dows}{Dowker,J.S. \aop{71}{1972}{577}; Dowker,J.S. and Pettengill,D.F.
  \jpa{7}{1974}{1527}; J.S.Dowker in {\it Quantum Gravity}, edited by
  S. C. Christensen (Hilger,Bristol,1984)}
  \ref{Jadhav2}{Jadhav,S.P. \prD{43}{1991}{2656}.}
  \ref{Dow11}{Dowker,J.S. \cqg{21}{2004}4247.}
  \ref{Dow12}{Dowker,J.S. \cqg{21}{2004}4977.}
  \ref{Dow13}{Dowker,J.S. \jpa{38}{2005}1049.}
  \ref{Zagier}{Zagier,D. \ma{202}{1973}{149}}
  \ref{RandG}{Rademacher, H. and Grosswald,E. {\it Dedekind Sums},
  (Carus, MAA, 1972).}
  \ref{Berndt7}{Berndt,B, \aim{23}{1977}{285}.}
  \ref{HKMM}{Harvey,J.A., Kutasov,D., Martinec,E.J. and Moore,G.
  {\it Localised Tachyons and RG Flows}, hep-th/0111154.}
  \ref{Beck}{Beck,M., {\it Dedekind Cotangent Sums}, {\it Acta Arithmetica}
  {\bf 109} (2003) 109-139 ; math.NT/0112077.}
  \ref{McInnes}{McInnes,B. {\it APS instability and the topology of the brane
  world}, hep-th/0401035.}
  \ref{BHS}{Brevik,I, Herikstad,R. and Skriudalen,S. {\it Entropy Bound for the
  TM Electromagnetic Field in the Half Einstein Universe}; hep-th/0508123.}
  \ref{BandO}{Brevik,I. and Owe,C.  \prD{55}{4689}{1997}.}
  \ref{Kenn}{Kennedy,G. Thesis. University of Manchester 1978.}
  \ref{KandU}{Kennedy,G. and Unwin S. \jpa{12}{L253}{1980}.}
  \ref{BandO1}{Bayin,S.S.and Ozcan,M.
  \prD{48}{2806}{1993}; \prD{49}{5313}{1994}.}
  \ref{Chang}{Chang, P. Thesis. University of Manchester 1993.}
  \ref{Barnesa}{Barnes,E.W. {\it Trans. Camb. Phil. Soc.} {\bf 19} (1903) 374.}
  \ref{Barnesb}{Barnes,E.W. {\it Trans. Camb. Phil. Soc.}
  {\bf 19} (1903) 426.}
  \ref{Stanley1}{Stanley,R.P. \joa {49Hilf}{1977}{134}.}
  \ref{Stanley}{Stanley,R.P. \bams {1}{1979}{475}.}
  \ref{Hurley}{Hurley,A.C. \pcps {47}{1951}{51}.}
  \ref{IandK}{Iwasaki,I. and Katase,K. {\it Proc.Japan Acad. Ser} {\bf A55}
  (1979) 141.}
  \ref{IandT}{Ikeda,A. and Taniguchi,Y. {\it Osaka J. Math.} {\bf 15} (1978)
  515.}
  \ref{GandM}{Gallot,S. and Meyer,D. \jmpa{54}{1975}{259}.}
  \ref{Flatto}{Flatto,L. {\it Enseign. Math.} {\bf 24} (1978) 237.}
  \ref{OandT}{Orlik,P and Terao,H. {\it Arrangements of Hyperplanes},
  Grundlehren der Math. Wiss. {\bf 300}, (Springer--Verlag, 1992).}
  \ref{Shepler}{Shepler,A.V. \joa{220}{1999}{314}.}
  \ref{SandT}{Solomon,L. and Terao,H. \cmh {73}{1998}{237}.}
  \ref{Vass}{Vassilevich, D.V. \plb {348}{1995}39.}
  \ref{Vass2}{Vassilevich, D.V. \jmp {36}{1995}3174.}
  \ref{CandH}{Camporesi,R. and Higuchi,A. {\it J.Geom. and Physics}
  {\bf 15} (1994) 57.}
  \ref{Solomon2}{Solomon,L. \tams{113}{1964}{274}.}
  \ref{Solomon}{Solomon,L. {\it Nagoya Math. J.} {\bf 22} (1963) 57.}
  \ref{Obukhov}{Obukhov,Yu.N. \pl{109B}{1982}{195}.}
  \ref{BGH}{Bernasconi,F., Graf,G.M. and Hasler,D. {\it The heat kernel
  expansion for the electromagnetic field in a cavity}; math-ph/0302035.}
  \ref{Baltes}{Baltes,H.P. \prA {6}{1972}{2252}.}
  \ref{BaandH}{Baltes.H.P and Hilf,E.R. {\it Spectra of Finite Systems}
  (Bibliographisches Institut, Mannheim, 1976).}
  \ref{Ray}{Ray,D.B. \aim{4}{1970}{109}.}
  \ref{Hirzebruch}{Hirzebruch,F. {\it Topological methods in algebraic
  geometry} (Springer-- Verlag,\break  Berlin, 1978). }
  \ref{BBG}{Bla\v{z}i\'c,N., Bokan,N. and Gilkey, P.B. {\it Ind.J.Pure and
  Appl.Math.} {\bf 23} (1992) 103.}
  \ref{WandWi}{Weck,N. and Witsch,K.J. {\it Math.Meth.Appl.Sci.} {\bf 17}
  (1994) 1017.}
  \ref{Norlund}{N\"orlund,N.E. \am{43}{1922}{121}.}
  \ref{Duff}{Duff,G.F.D. \aom{56}{1952}{115}.}
  \ref{DandS}{Duff,G.F.D. and Spencer,D.C. \aom{45}{1951}{128}.}
  \ref{BGM}{Berger, M., Gauduchon, P. and Mazet, E. {\it Lect.Notes.Math.}
  {\bf 194} (1971) 1. }
  \ref{Patodi}{Patodi,V.K. \jdg{5}{1971}{233}.}
  \ref{GandS}{G\"unther,P. and Schimming,R. \jdg{12}{1977}{599}.}
  \ref{MandS}{McKean,H.P. and Singer,I.M. \jdg{1}{1967}{43}.}
  \ref{Conner}{Conner,P.E. {\it Mem.Am.Math.Soc.} {\bf 20} (1956).}
  \ref{Gilkey2}{Gilkey,P.B. \aim {15}{1975}{334}.}
  \ref{MandP}{Moss,I.G. and Poletti,S.J. \plb{333}{1994}{326}.}
  \ref{BKD}{Bordag,M., Kirsten,K. and Dowker,J.S. \cmp{182}{1996}{371}.}
  \ref{RandO}{Rubin,M.A. and Ordonez,C. \jmp{25}{1984}{2888}.}
  \ref{BaandD}{Balian,R. and Duplantier,B. \aop {112}{1978}{165}.}
  \ref{Kennedy2}{Kennedy,G. \aop{138}{1982}{353}.}
  \ref{DandKi2}{Dowker,J.S. and Kirsten, K. {\it Analysis and Appl.}
 {\bf 3} (2005) 45.}
  \ref{Dow40}{Dowker,J.S. {\it p-form spectra and Casimir energy}
  hep-th/0510248.}
  \ref{BandHe}{Br\"uning,J. and Heintze,E. {\it Duke Math.J.} {\bf 51} (1984)
   959.}
  \ref{Dowl}{Dowker,J.S. {\it Functional determinants on M\"obius corners};
    Proceedings, `Quantum field theory under
    the influence of external conditions', 111-121,Leipzig 1995.}
  \ref{Dowqg}{Dowker,J.S. in {\it Quantum Gravity}, edited by
  S. C. Christensen (Hilger, Bristol, 1984).}
  \ref{Dowit}{Dowker,J.S. \jpa{11}{1978}{347}.}
  \ref{Kane}{Kane,R. {\it Reflection Groups and Invariant Theory} (Springer,
  New York, 2001).}
  \ref{Sturmfels}{Sturmfels,B. {\it Algorithms in Invariant Theory}
  (Springer, Vienna, 1993).}
  \ref{Bourbaki}{Bourbaki,N. {\it Groupes et Alg\`ebres de Lie}  Chap.III, IV
  (Hermann, Paris, 1968).}
  \ref{SandTy}{Schwarz,A.S. and Tyupkin, Yu.S. \np{242}{1984}{436}.}
  \ref{Reuter}{Reuter,M. \prD{37}{1988}{1456}.}
  \ref{EGH}{Eguchi,T. Gilkey,P.B. and Hanson,A.J. \prp{66}{1980}{213}.}
  \ref{DandCh}{Dowker,J.S. and Chang,Peter, \prD{46}{1992}{3458}.}
  \ref{APS}{Atiyah M., Patodi and Singer,I.\mpcps{77}{1975}{43}.}
  \ref{Donnelly}{Donnelly.H. {\it Indiana U. Math.J.} {\bf 27} (1978) 889.}
  \ref{Katase}{Katase,K. {\it Proc.Jap.Acad.} {\bf 57} (1981) 233.}
  \ref{Gilkey3}{Gilkey,P.B.\invm{76}{1984}{309}.}
  \ref{Degeratu}{Degeratu.A. {\it Eta--Invariants and Molien Series for
  Unimodular Groups}, Thesis MIT, 2001.}
  \ref{Seeley}{Seeley,R. \ijmp {A\bf18}{2003}{2197}.}
  \ref{Seeley2}{Seeley,R. .}
  \ref{melrose}{Melrose}
  \ref{berard}{B\'erard,P.}
  \ref{gromes}{Gromes,D.}
  \ref{Ivrii}{Ivrii}
  \ref{DandW}{Douglas,R.G. and Wojciekowski,K.P. \cmp{142}{1991}{139}.}
  \ref{Dai}{Dai,X. \tams{354}{2001}{107}.}
  \ref{Kuznecov}{Kuznecov}
  \ref{DandG}{Duistermaat and Guillemin.}
  \ref{PTL}{Pham The Lai}
\end{putreferences}

\bye